\documentclass[useAMS,usenatbib]{mn2e}

\usepackage{dsfont}
\usepackage{amsmath}
\usepackage{bm}
\usepackage{amssymb}
\usepackage{graphicx}
\usepackage{verbatim}
\usepackage{natbib}
\usepackage{eucal}
\usepackage{calligra}

\usepackage{amsfonts}
\usepackage{color}
\usepackage[normalem]{ulem}
\usepackage[T1]{fontenc}

\usepackage{times,epstopdf}
\usepackage{epsfig}
\usepackage[usenames,dvipsnames]{xcolor}
\usepackage{url}
\usepackage{subfigure}
\DeclareMathAlphabet{\mathscr}{OT1}{pzc}%
                                 {m}{it}
                                 
\newcommand{\mnras}{MNRAS}
\newcommand{\jcap}{JCAP}
\newcommand{\apj}{ApJ}
\newcommand{\prd}{Phys. Rev. D}
\newcommand{\nat}{Nature (London)}

\newcommand{\be}{\begin{equation}}
\newcommand{\ee}{\end{equation}}
\newcommand{\bes}{\begin{equation*}}
\newcommand{\ees}{\end{equation*}}
\newcommand{\bea}{\begin{eqnarray}}
\newcommand{\eea}{\end{eqnarray}}
\newcommand{\beas}{\begin{eqnarray*}}
\newcommand{\eeas}{\end{eqnarray*}}

\newcommand{\Mpc}{\,h^{-1}{\rm Mpc}}
\newcommand{\Gpc}{\,h^{-1}{\rm Gpc}}
\newcommand{\Msun}{\,h^{-1}{\rm M_{\odot}}}
\def\apj{{ApJ}}                 
\def\apjl{{ApJ}}                
\def\apjs{{ApJS}}               
\def\ao{{Appl.~Opt.}}           
\def\aap{{A\&A}}                
\def\mnras{{MNRAS}}             
\def\prd{{Phys.~Rev.~D}}        
\def\pasj{{PASJ}}               
\def\nat{{Nature}}              

\begin{document} 

\title[Void RSD]
{Redshift-space distortions around voids}
\author[Cai et al.]
{Yan-Chuan Cai \thanks{E-Mail: cai@roe.ac.uk}$^1$, Andy Taylor$^1$, John A. Peacock$^1$ and Nelson Padilla$^{2,3}$ \\
$^{1}$Institute for Astronomy, University of Edinburgh, Royal Observatory, Edinburgh EH9 3HJ, UK\\
$^2$Instituto de Astrof\'isica, Pontificia Universidad Cat\'olica de Chile, Av. Vicu\~na Mackenna 4860, Santiago, Chile. \\                    
$^3$Centro de Astro-Ingenier\'ia, Pontificia Universidad Cat\'olica de Chile, Av. Vicu\~na Mackenna 4860, Santiago, Chile. }
\maketitle

\begin{abstract}
We have derived estimators for the linear growth rate of density fluctuations using the cross-correlation function (CCF) of voids and haloes in redshift space. In linear theory, this CCF contains only monopole and quadrupole terms. At scales greater than the void radius, linear theory is a good match to voids traced out by haloes; small-scale random velocities are unimportant at these radii, only tending to cause small and often negligible elongation of the CCF near its origin. By extracting the monopole and quadrupole from the CCF, we measure the linear growth rate without prior knowledge of the void profile or velocity dispersion. We recover the linear growth parameter $\beta$ to 9\% precision from an effective volume of $3(\Gpc)^3$ using voids with radius >25$\Mpc$. Smaller voids are predominantly sub-voids, which may be more sensitive to the random velocity dispersion; they introduce noise and do not help to improve measurements. Adding velocity dispersion as a free parameter allows us to use information at radii as small as half of the void radius. The precision on $\beta$ is reduced to 5\%. Voids show diverse shapes in redshift space, and can appear either elongated or flattened along the line of sight. This can be explained by the competing amplitudes of the local density contrast, plus the radial velocity profile and its gradient. The distortion pattern is therefore determined solely by the void profile and is different for void-in-cloud and void-in-void. This diversity of redshift-space void morphology complicates measurements of the Alcock-Paczynski effect using voids.
\end{abstract}


\section{Introduction}
Using redshift-space distortions (RSD) to probe the growth of
large-scale structure has been a target for cosmological research
since the first prediction of the effect \citep{Kaiser1987} and
observational RSD studies have been pursued for over two decades
\citep[e.g.][]{Hamilton1992, Cole1994, Cole1995, Peacock2001,
  Beutler2012, Reid2012,Blake2012, delaTorre2013, Samushia2014,
  Howlett2015, Okumura2015b}. The pairwise galaxy-galaxy power
spectrum or correlation function approaches the prediction of linear
theory at very large scales, but in the quasi-linear and non-linear
regime, more sophisticated models are needed to account for non-linear
growth \citep[e.g.][]{Scoccimarro2004, Matsubara2008,Taruya2009,
  Percival2009, Taruya2010,Seljak2011,Jennings2011,
  Gil-Marin2012,Okumura2012, delaTorre2012,Valageas2013,Okumura2015a};
with care, it is possible to recover the linear growth rate to 2\%
precision by including this small-scale information \citep{Reid2014}. This
{\it average\/} growth is extracted from the pairwise galaxy-galaxy
correlations, which sample peaks and troughs of the matter density
field; it remains an open question, at least from the observational
point of view, how the growth of structure depends on the
environment. At a minimum, the growth of density perturbations is
expected to be more rapid in superclusters and lower in voids, simply
because these regions resemble universes with different cosmological
parameters.  The scales where such non-linear growth effects become
important will probably differ between different environments.  But
the environmental dependence of the growth of structure may have a
more fundamental significance, since it could encode information about
non-standard theories of gravity. Measurements for the growth at
different environment can thus be used as a test for departures from
Einstein gravity.  As discussed below, such deviations are frequently
expected to be stronger where the matter density is low, which leads
us to investigate the growth of structure in void environments.

Cosmic voids are large underdense regions of the universe that are
devoid of galaxies. Voids in large-scale structure have great
potential in constraining cosmology and gravity via the following
observables: the Alcock-Paczynski (AP) test \citep{AP1979,LavauxWandelt11};
stacking of voids for the integrated Sachs-Wolfe (ISW) effect 
\citep{Sachs1967, Granett2008,Cai2014a,Cai2014b}; 
weak lensing measurement of the matter distribution in
voids \citep{Higuchi2013, Krause2013, Cai2015, Clampitt2015,
  Gruen2015}; void ellipticity as a probe for the dark energy equation
of state \citep{Lee2009, Bos2012}; void density profiles and number counts as a
probe of modified gravity \citep{Clampitt2012, Lam2015, Cai2015, Zivick2015,
  Barreira2015}; coupled dark energy \citep{Pollina2015}; the
nature of dark matter \citep{Yang2015}; massive neutrino \citep{Massara2015}; and Baryon Acoustic Oscillations in void clustering \citep{Kitaura2015, Liang2015}.
 
The growth of structure around cosmic voids is fundamentally related
to the detailed understanding of some of these observables.  For
example, the AP measurement using cosmic voids makes the underlying
assumption that stacked voids are of the same size in both the
transverse and line-of-sight (LOS) direction if the assumed cosmology
is correct -- but this assumption is violated by redshift-space
distortions \citep{LavauxWandelt11, Sutter2014}. The effect of peculiar velocities
on the observed
configuration of voids must be
understood in order to obtain unbiased cosmological AP measurements.
Another example is that the ISW signal associated with
cosmic voids is determined by the local growth rate, which is affected
by possibly nonlinear density and velocity structure.

We will focus here on extracting the growth rate around voids.  The
void-mass (or void-galaxy) correlations in redshift space are the tool
we will employ for the measurement. We will follow closely the
methodology of conventional redshift-space distortion analyses to
derive the mapping between redshift-space and real-space clustering
for voids. This will give us a comprehensive picture of how the
redshift-space void-mass correlation function is affected by different
aspects of the density and velocities, which will be essential for
understanding how the AP measurement with voids is affected by
redshift-space distortions.

In Section 2, we derive the expression for the redshift-space
void-mass correlation function using linear theory.
In Section 3 we then give
examples to explain the complexity of void-mass correlation
function. Section 4 focuses on developing a method for measuring the
growth around voids and testing it using an N-body simulation.  During the
preparation of this manuscript, \cite{Hamaus2015} released a paper on
the same topic; we compare our results at the end of Section 4. We sum
up and draw conclusions in Section 5.
\section{Redshift-space distortions around voids}
Although our goal is to derive the expression for the void-mass correlation function in redshift space in this section, there is no requirement that the system must be underdense. The derivation and the results are therefore relevant to the overdense case, i.e. to the halo-mass correlation function -- see e.g. \cite{Croft1999, ZuWeinberg2013}.

\subsection{Linear theory}
The difference between the redshift-space void-mass correlation
function and the pairwise galaxy-galaxy correlation function is that
we are considering the relative peculiar velocities of dark matter (or
dark matter haloes) with respect to one central point, the void
centre. The bulk motion of voids will therefore not affect the
void-mass correlation function within the scales where the
bulk velocity field can be considered to be coherent (see 
Section~\ref{Sec:Measurement} for further discussion on this point).
With the plane-parallel approximation,
$|v/r_c aH|\ll 1$ where $v$ is the peculiar velocity of dark matter,
$r_c$ is its comoving distance from the observer, $a$ is the scale
factor of the universe and $H$ is the Hubble constant at $a$, the
mapping between redshift-space and real-space overdensities is
\begin{equation}
1+\delta^s({\bf r})=[1+ \delta({\bf r})]\; [1+ u'({\bf r})]^{-1}, 
\end{equation}
where $\delta({\bf r})=\rho({\bf r})/\bar{\rho}-1$ with $\bar{\rho}$
being the mean density of the universe and $\rho({\bf r})$ the matter
density at ${\bf r}$, which is defined with an origin at the void centre.
For $\delta({\bf r})\ll 1$ and $|\partial v ({\bf r})/\partial {\bf
  r}| \ll aH$, to linear order we have
\begin{equation}
\label{Eq:2}
\delta^s({\bf r})=\delta({\bf r})- u'({\bf r}).
\end{equation}
The distortion term $u'({\bf r})$ is the gradient of the radial velocity profile around the void centre $v({\bf r})$
projected along the line of sight:
\begin{align}
u'(\mathbf r)&={\bf \hat z} \cdot \frac{1}{aH}\frac{\partial \left[{\bf \hat z \cdot \hat r} v({\bf r})\right] }{\partial {\bf r}}\\
\label{Eq:4}
&= (1-\mu^2)\frac{1}{aH}\frac{v({ r})}{r}+\mu^2\frac{1}{aH}\frac{\partial v( {r})}{\partial r},
\end{align}
where ${\bf \hat z}$ is the unit vector along 
the line of sight, $\mu=\cos(\theta)$ and $\theta$ is the angle subtended between ${\bf r}$ and 
${\bf \hat z}$. In linear theory,
\begin{align}
v(r)&= -\frac{1}{3} raH \bar{\delta}(r) f; \\
\label{Eq:xibar}
\bar{\delta}(r)&=\frac{3}{r^3}\int^r_0\delta(r')r'^2dr'
\end{align}
\citep{Peebles1993}, where $f \equiv d \ln D /d \ln a$ is the linear growth rate, 
and $D$ is the linear growth factor.  It is straightforward to show that 
\begin{equation}
\frac{\partial v(r)}{\partial r}=-faH\left[\delta(r)-\frac{2}{3}\bar{\delta}(r)\right].
\end{equation}
Eq.~(\ref{Eq:2}) then takes the form
\begin{equation}
\label{Eq:zspaceProfile}
\xi_{\rm vm}^s({\bf r})=\delta(r)+\frac{1}{3}f\bar{\delta}(r)+f\mu^2[\delta(r)-\bar{\delta}(r)], 
\end{equation}
where instead of $\delta^s({\bf r})$, we have used the notation
$\xi_{\rm vm}^s({\bf r})$. This is because the conditional
density fluctuation is identical in meaning to the cross-correlation
function, which gives the fractional fluctuation in the number
of cross-pairs.
The above equation therefore shows that
the void-mass correlation function in redshift space contains only
monopole and quadrupole terms. This is different from the pairwise
galaxy-galaxy autocorrelation function, where there is also a
hexadecapole moment.  We discuss the reason for this distinction in
greater depth in Section~\ref{Sec:Measurement}.

\subsection{Quasi-linear model}
\label{Quasi}
The above expression is only valid on the assumption that $\delta \ll
1$ and that any random dispersion in velocity is small. These assumptions can be
relaxed, which leads to the quasi-linear model \citep{Peebles1993,
  Fisher1994a, Fisher1994b, Fisher1995}. In this model, the redshift-space 
correlation function is expressed as the convolution of its real-space 
version with the probability distribution function for
velocities along the LOS:
\begin{align}
1+\xi_{\rm vm}^s(r_{\sigma},r_{\pi})=&
\int_{-\infty}^{\infty}\left[1+\delta\left(r_{\sigma}, 
 r_{\pi}-\frac{v_{\pi}}{aH}\right)\right]\times p(v_{\pi})dv_{\pi} \nonumber \\ 
 =&\int_{-\infty}^{\infty}\left[1+\delta\left(r_{\sigma},
  r_{\pi}-\frac{v_{\pi}}{aH}\right)\right]\times \nonumber \\ 
&\frac{1}{\sqrt{2\pi}\sigma_{\rm v} } \exp\left(-\frac{\left[(v_{\pi}-v(r)\mu\right]^2}{2\sigma_{\rm v}^2(r)}\right)
dv_{\pi},
\label{FisherModel}
\end{align}
where $\mu=(r_{\pi}-v_{\pi}/aH ) / r$,
$r^2=r_{\sigma}^2+(r_{\pi}-v_{\pi}/aH)^2$. When $\delta \ll 1$ and
$\sigma_{\rm v} (r)$ is small, linear theory is well recovered
\citep{Peebles1993, Fisher1995}, as we have also explicitly checked.

With the assumption that galaxies or haloes are linearly biased
tracers of dark matter, we can generalise this result to give the
void-galaxy or void-halo correlation function by replacing $f$ with
$\beta$ in the above equation, where $\beta=f/b$, and $b$ is the
linear bias for galaxies or haloes.

\begin{figure*}
\begin{center}
\scalebox{0.5}{
\includegraphics[angle=0]{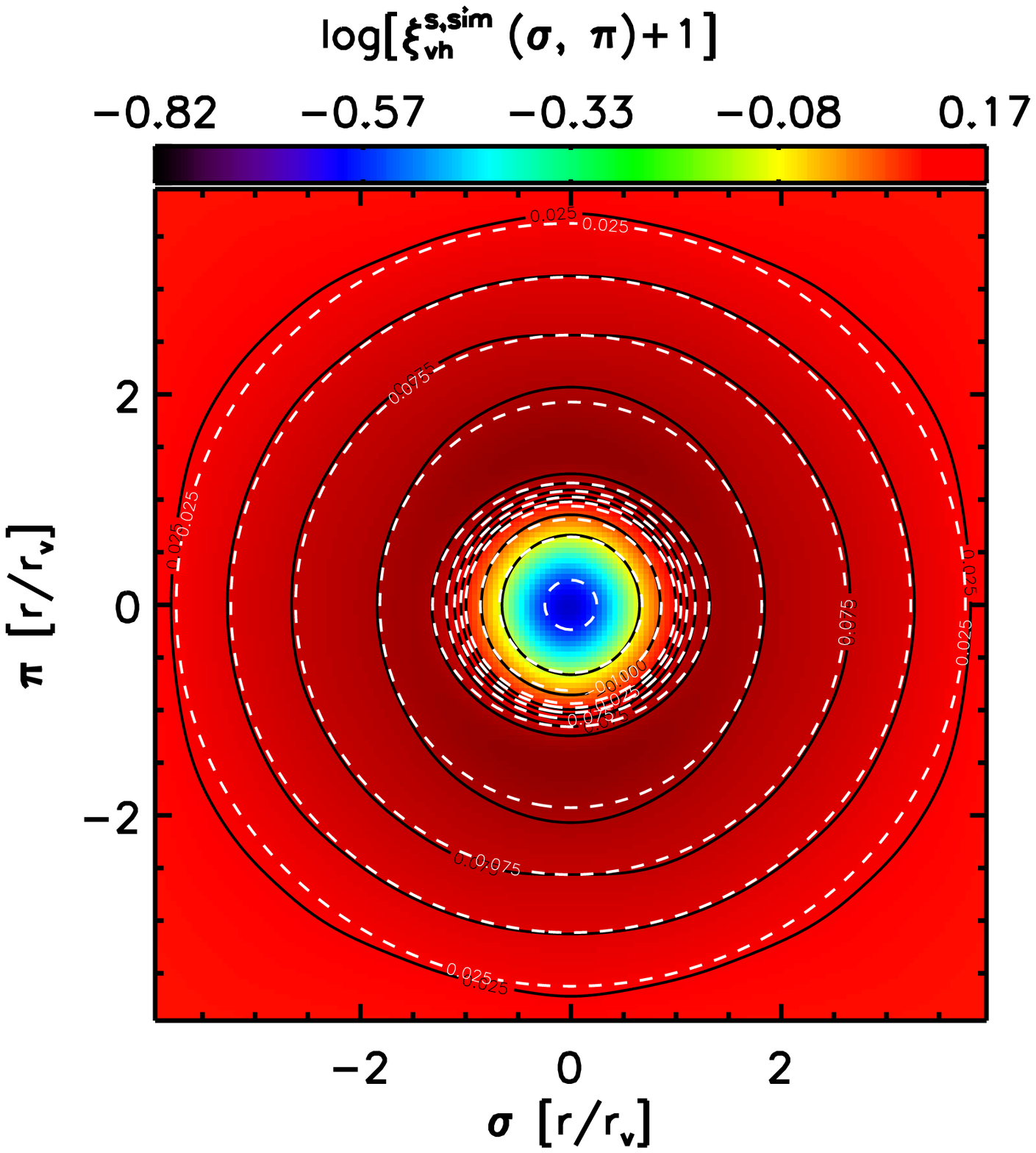}
\hspace{-3 cm}
\includegraphics[angle=0]{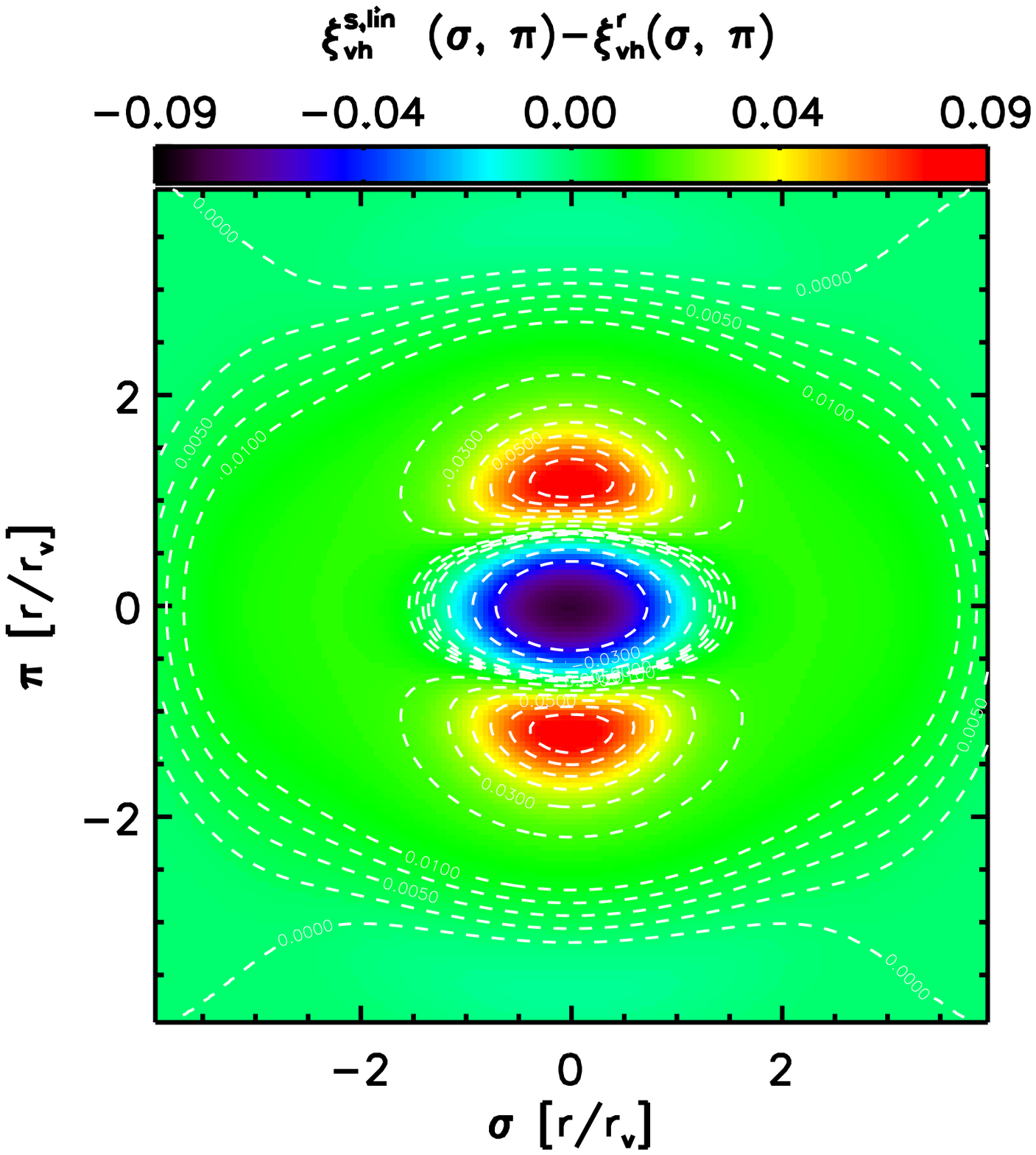}}
\scalebox{0.46}{
\hspace{-2.7 cm}
\includegraphics[angle=0]{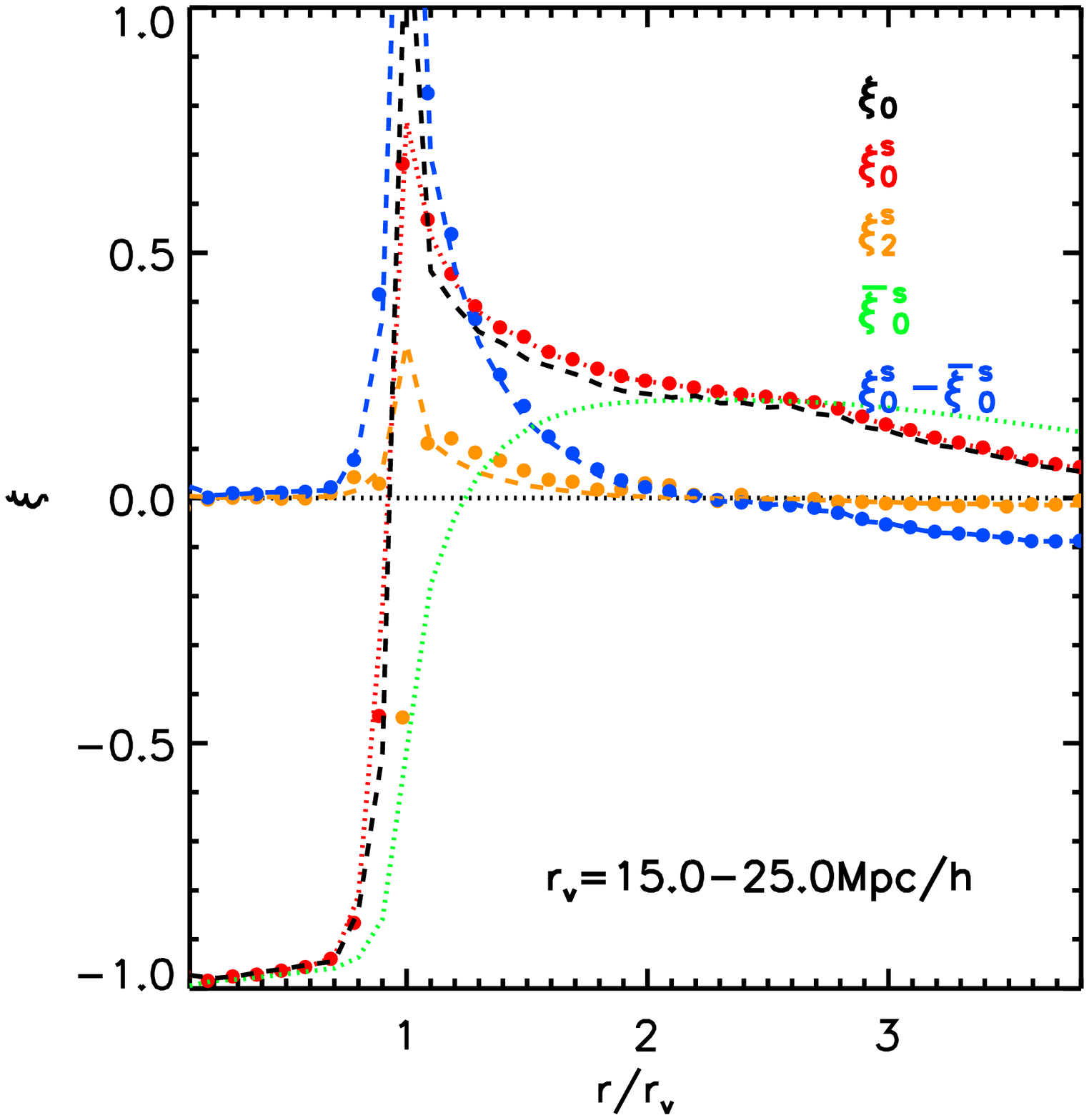}
\hspace{-1.5 cm}
\includegraphics[angle=0]{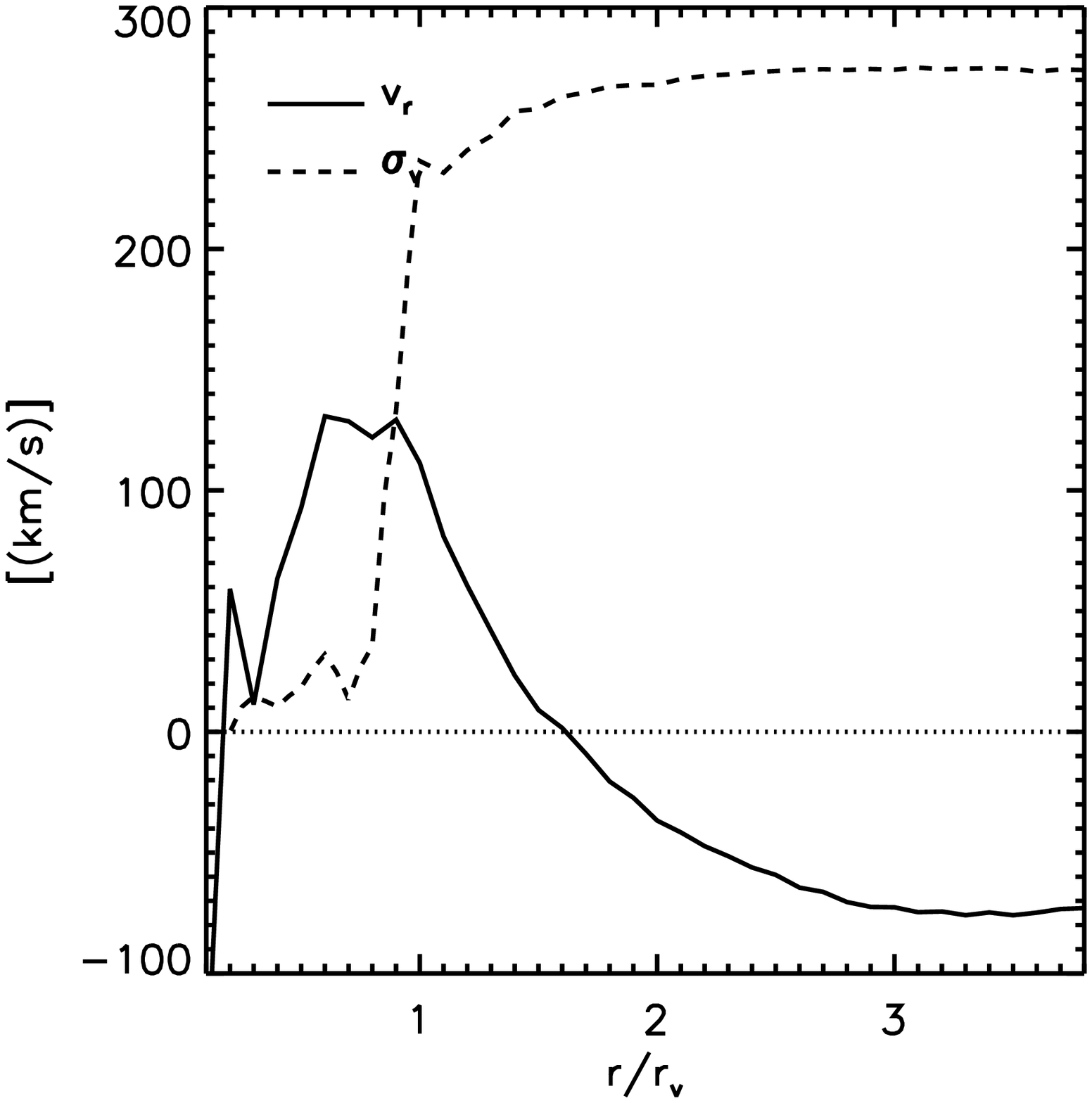}}
\caption{Top-left: redshift-space void-halo correlation functions. In
  this example, only voids-in-clouds defined as
  $\bar{\xi}^s_0(r=3r_{\rm v})>0$ are selected from the sample.  Black
  contours and the underlying colour image are results from our
  simulation. White-dashed contours are from the best-fit linear
  model. Top-right: The distortions to the correlation function due to
  peculiar velocities from the linear model. This is obtained by
  subtracting the real-space correlation function from its
  redshift-space version. Bottom-left: Moments of the real and
  redshift-space void-halo correlation functions. $\xi_0$ shown by the
  black dashed curve is the real-space monopole measured from
  our simulation using 3D positions of haloes in real space. It is in
  essence the stacked void profile in real space. The red-dashed curve
  ($\xi_0^s$ ) is the redshift-space void profile measured from the 3D
  positions of haloes in redshift space. The red dots are the same as
  the red-dashed curve except that they are the monopoles extracted
  from the correlation function shown on the top-left panel. The
  orange (quadrupole $\xi_2^s$) and blue dots
  ($\xi_0^s-\bar{\xi}^s_0$) are measurements from the correlation
  function, and the dashed curves of the same colours are predictions
  from linear theory. The green dotted curve is $\bar{\xi}^s_0$, the
  cumulative void profile. Bottom-right: radial velocity profile from
  void centres (solid black) and its dispersion (black dashed
  curve).}
\label{Fig:LinearModel1}
\end{center}
\end{figure*}

\begin{figure*}
\begin{center}
\scalebox{0.5}{
\includegraphics[angle=0]{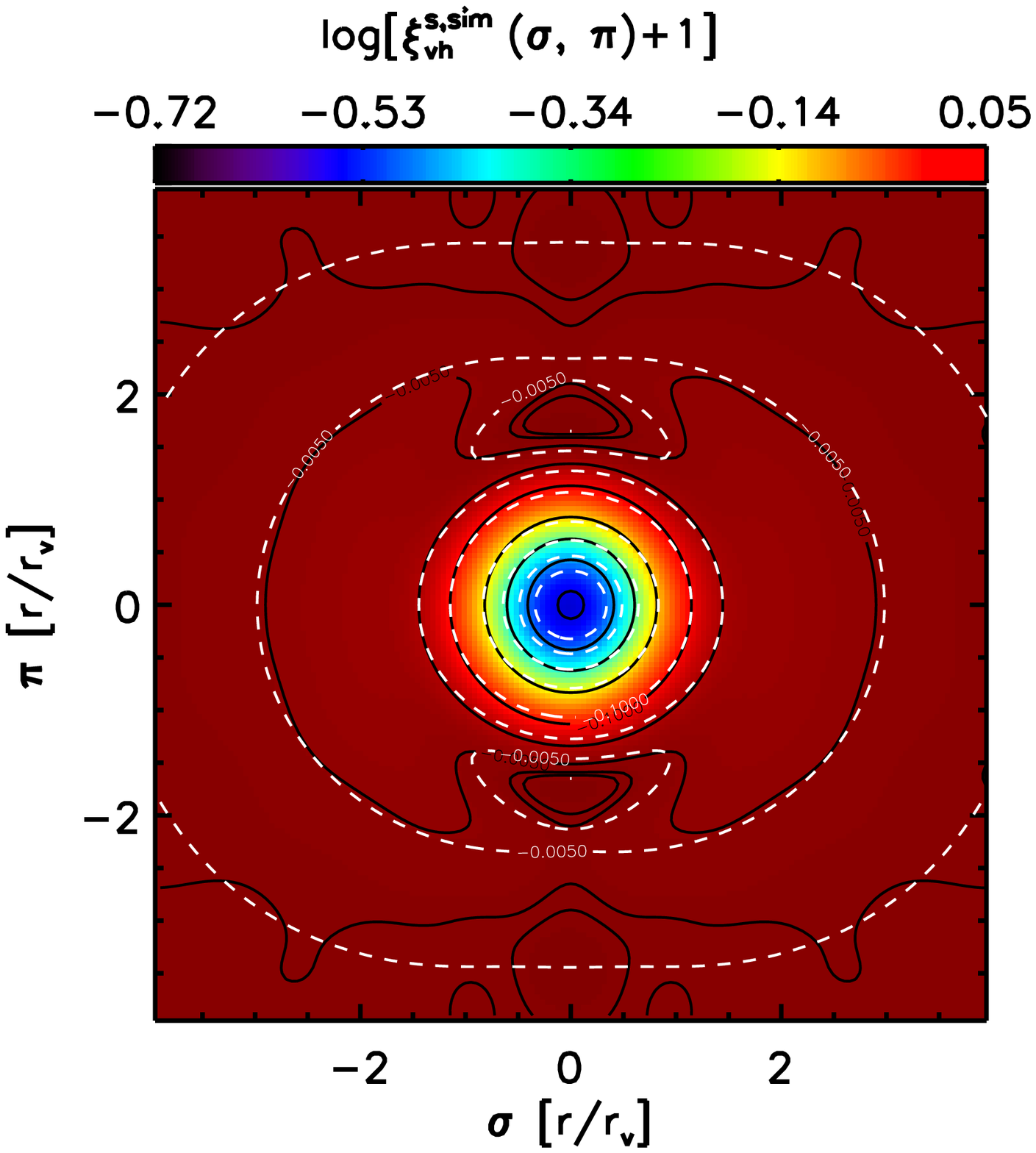}
\hspace{-3 cm}
\includegraphics[angle=0]{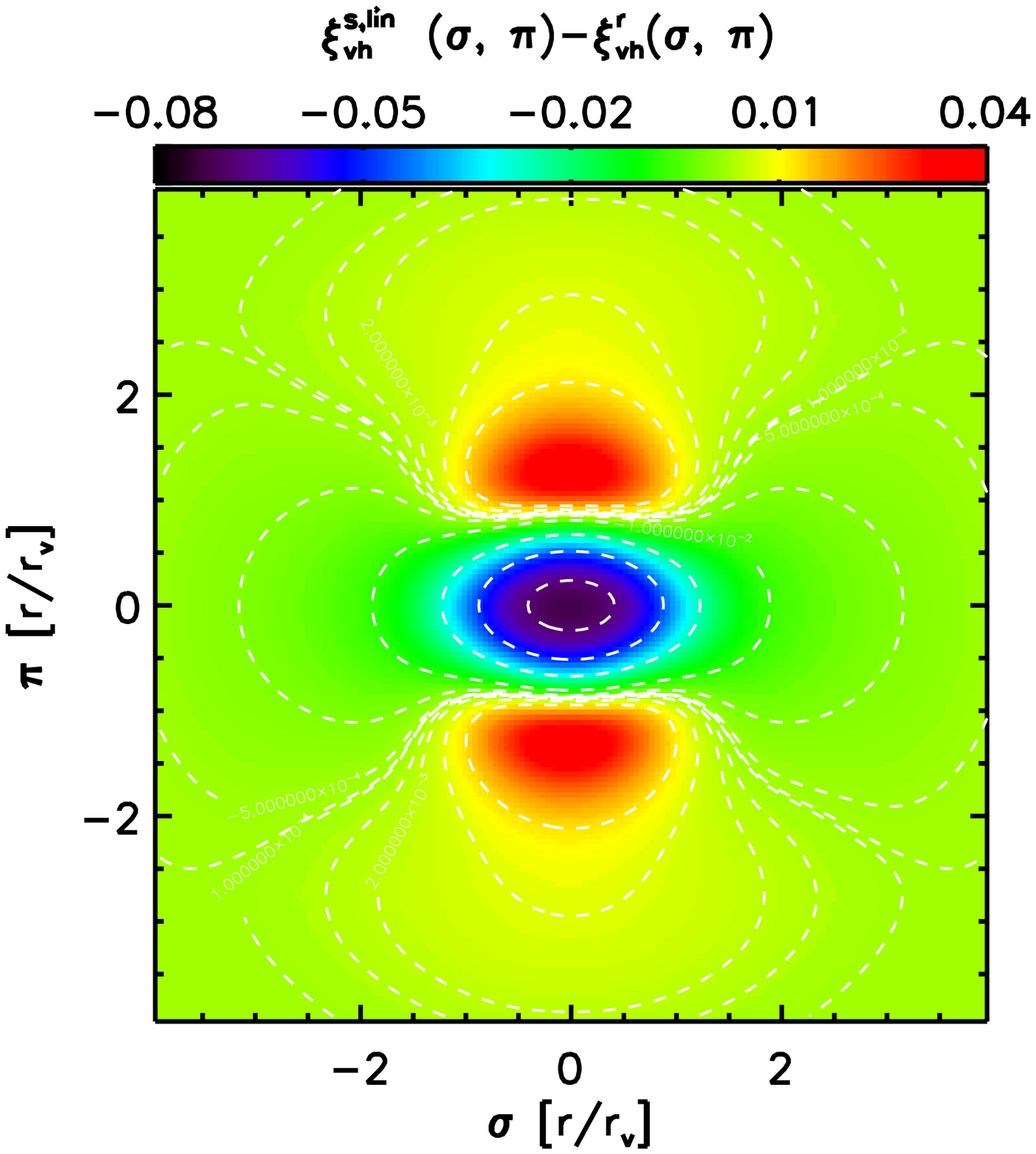}}
\scalebox{0.46}{
\hspace{-2.7 cm}
\includegraphics[angle=0]{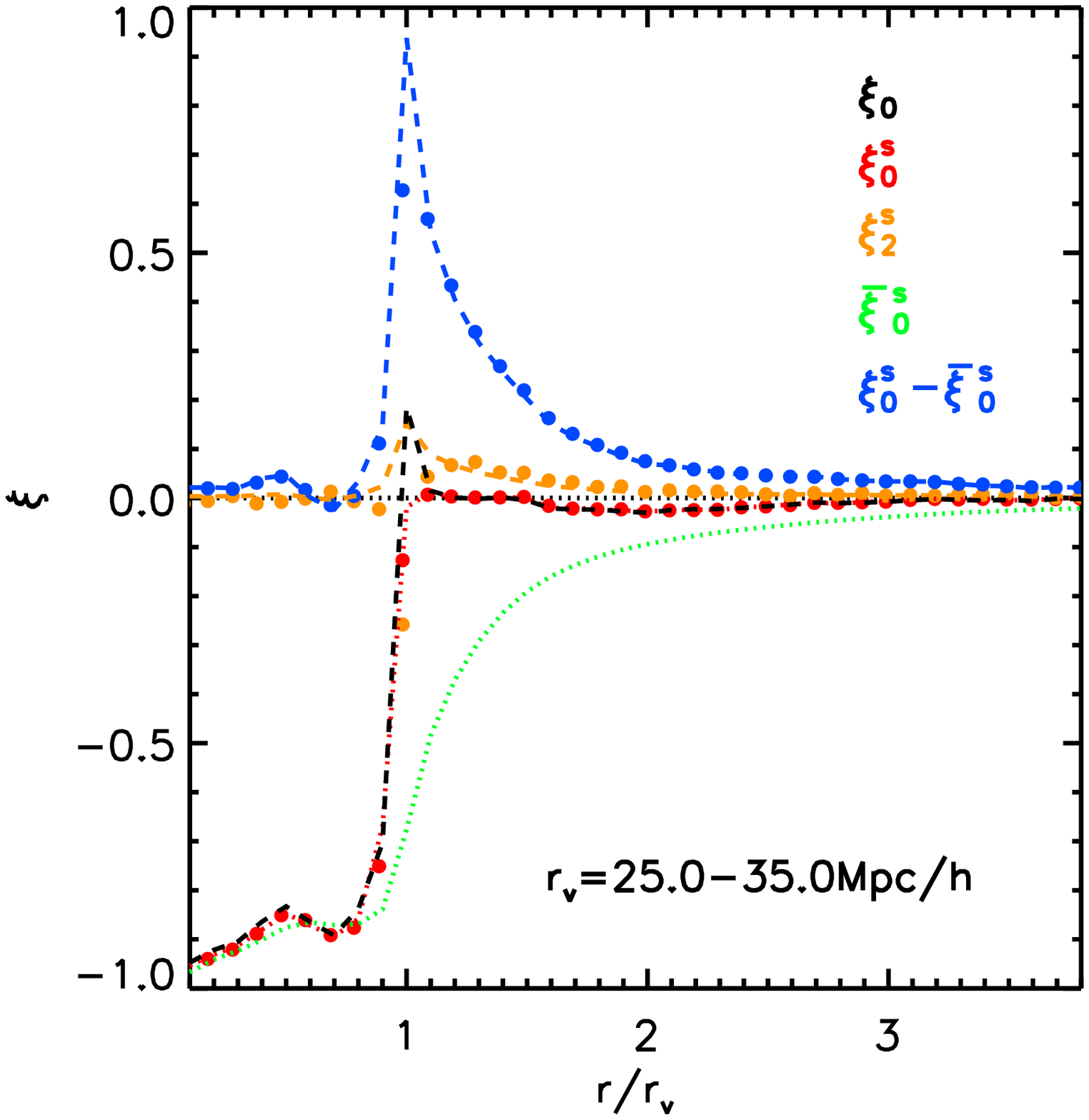}
\hspace{-1.5 cm}
\includegraphics[angle=0]{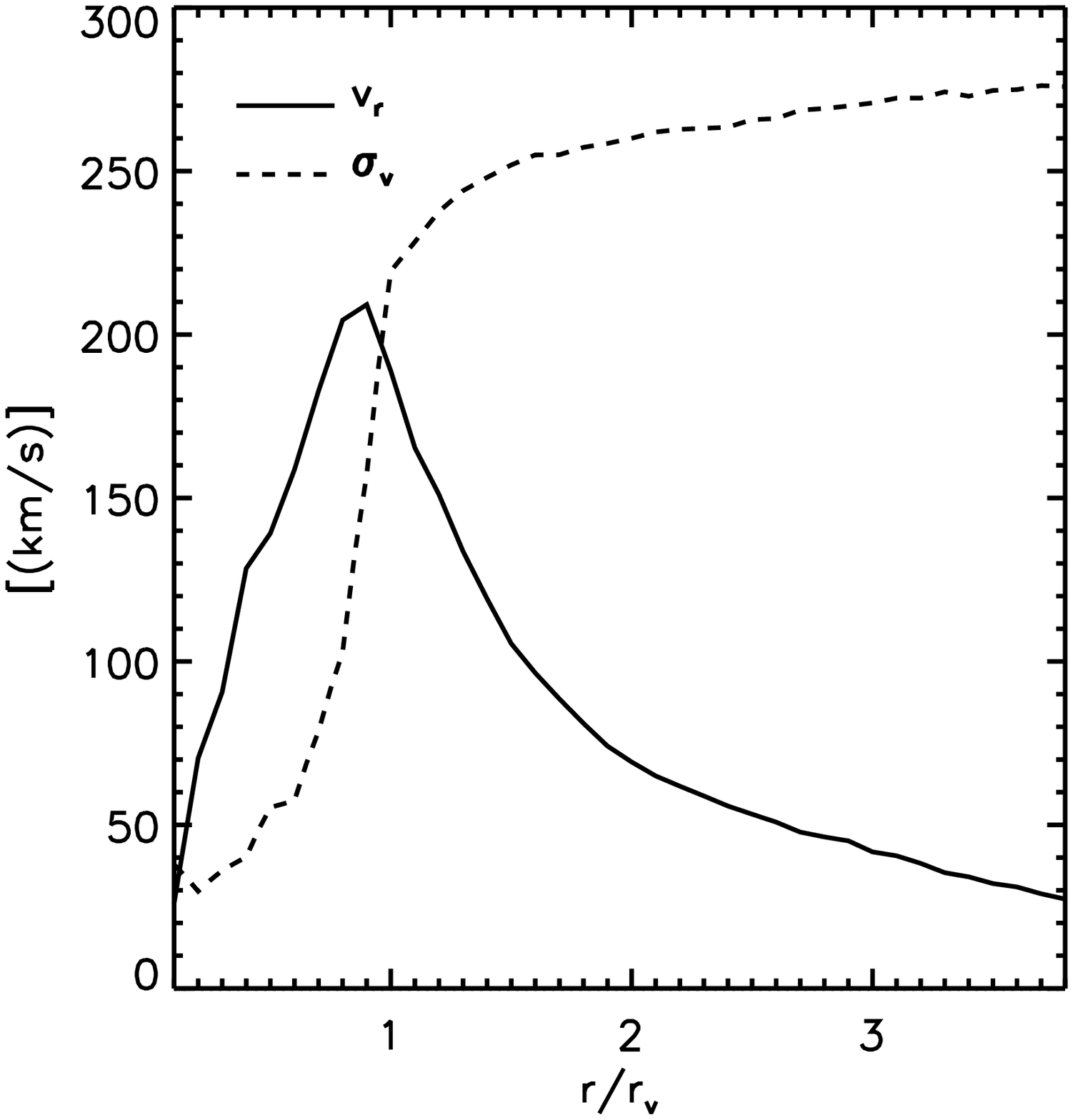}}
\caption{The same as Fig.~\ref{Fig:LinearModel1} but showing all voids with radii between 25-35$\Mpc$.}
\label{Fig:LinearModel2}
\end{center}
\end{figure*}

\begin{figure*}
\begin{center}
\scalebox{0.5}{
\includegraphics[angle=0]{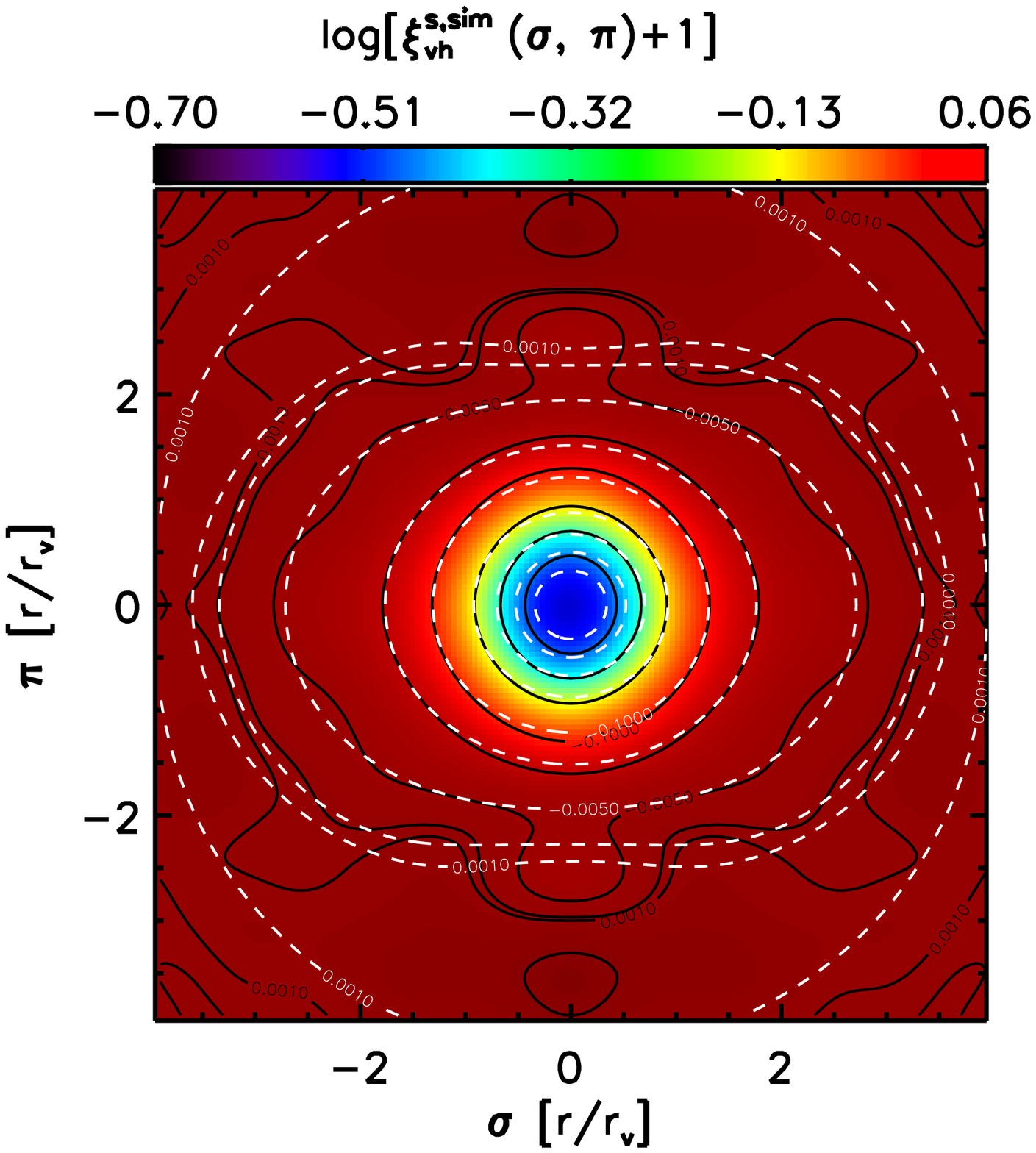}
\hspace{-3 cm}
\includegraphics[angle=0]{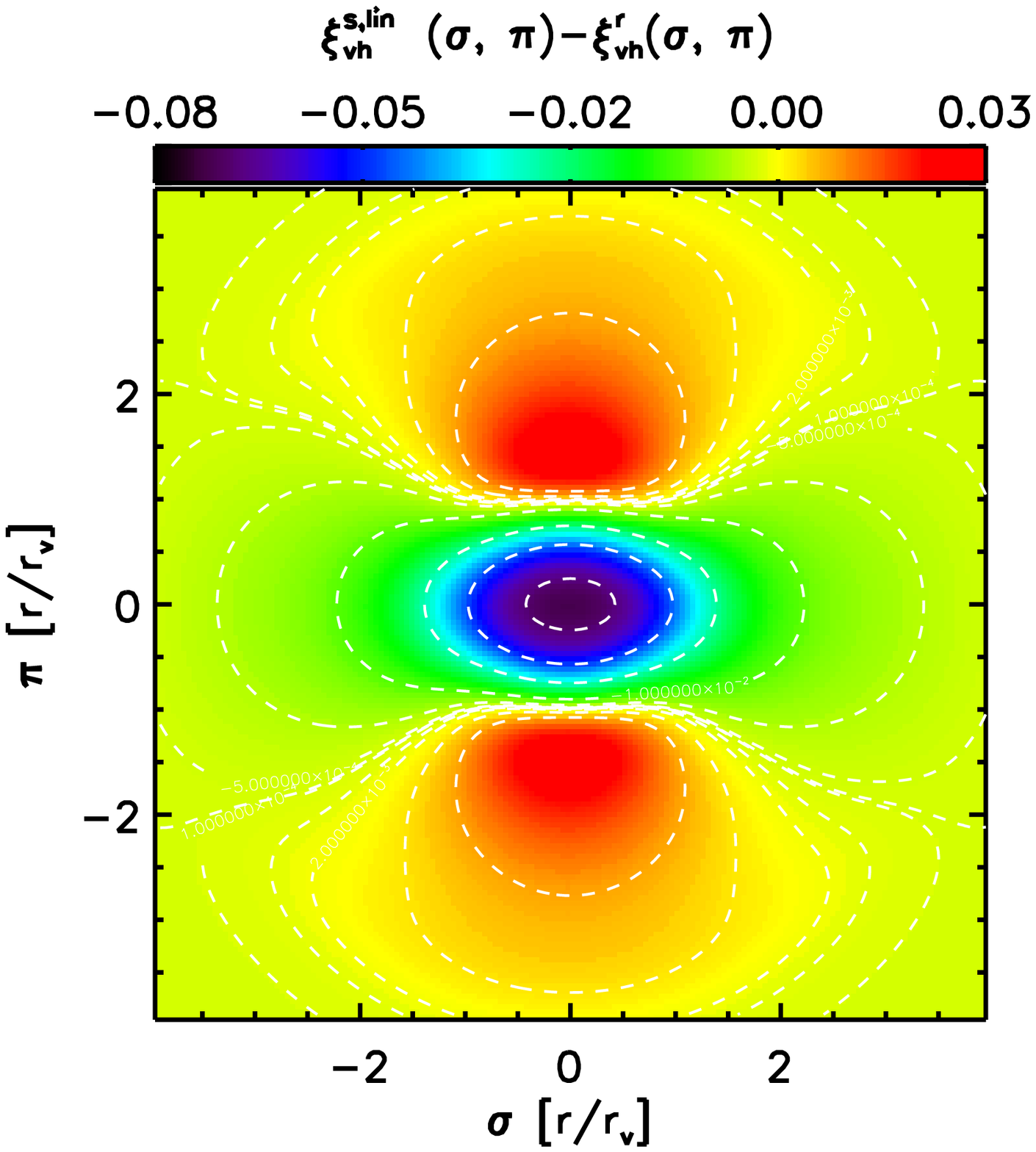}}
\scalebox{0.46}{
\hspace{-2.7 cm}
\includegraphics[angle=0]{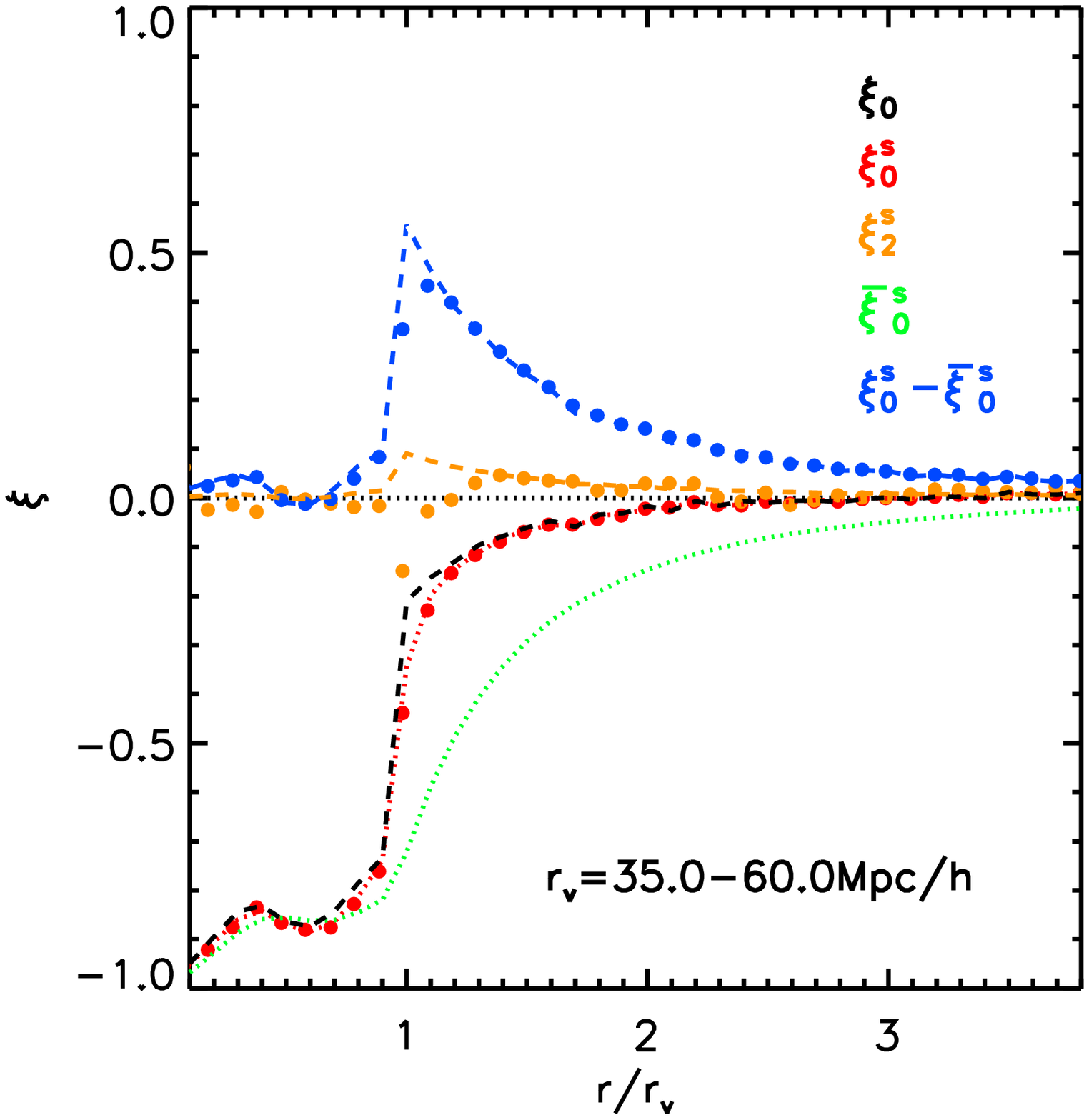}
\hspace{-1.5 cm}
\includegraphics[angle=0]{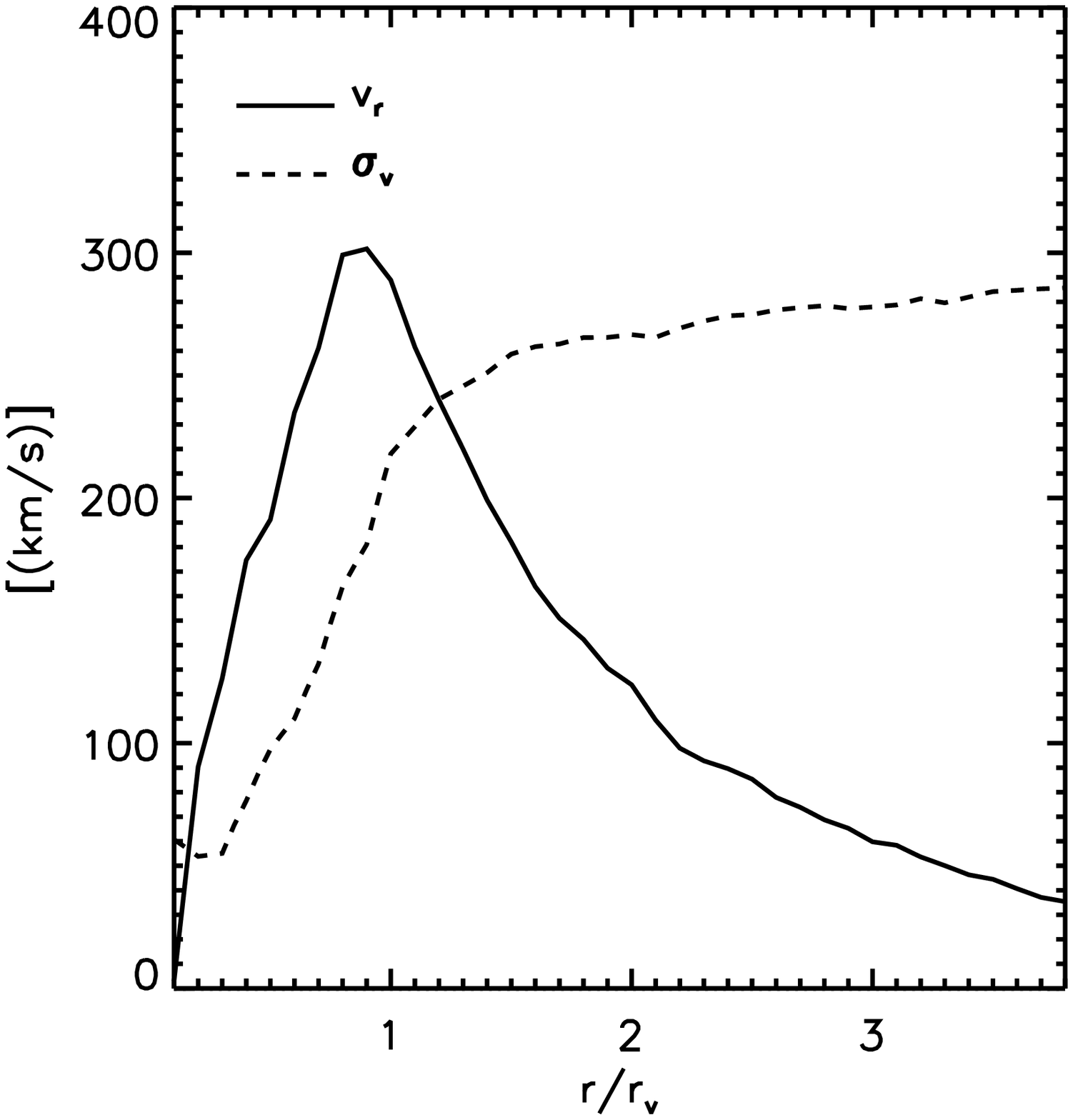}}
\caption{The same as Fig.~\ref{Fig:LinearModel1} but showing all voids with radii between 35-60$\Mpc$.}
\label{Fig:LinearModel3}
\end{center}
\end{figure*}
\subsection{Simulation set up and void definition}
We employ an N-body simulation of a $\Lambda$CDM model to test the performance of 
the linear model and later to test our estimator to extract the linear growth from voids.
The simulation was run with the
following parameters: $\Omega_{\rm m}=0.24$, $\Omega_{\rm
  \Lambda}=0.76$, $h = 0.73$, and $n_{\rm s}=0.958$ and
$\sigma_8=0.80$ \citep{Li2013}. The volume of the simulation box is
$(1\Gpc)^3$ We use all haloes above a minimum halo mass of
$M_{\rm min}=10^{12.8} \Msun$ to ensure that each halo contains at
least 100 particles. We choose to test our analysis at $z=0$, where our halo number density is 3.1$\times10^{-4}(\Mpc)^{-3}$, and the linear 
halo bias of our halo population is $b=1.7$. Voids are found in the halo field with the
spherical underdensity algorithm described in \cite{Cai2015}, which
is based on the algorithm of \cite{Padilla2005}.  The algorithm works
on the halo catalogue in the simulation box by growing maximal spheres
from a set of grid points within which the number density of haloes
satisfies the criterion $\Delta \le 0.2$. Void candidates are ranked
in decreasing order of radius.  Spheres that overlap with a neighbour
by a radius more than 50\% of the sum of their radii are
rejected. Note that this does not mean that no sub-voids are
allowed. A small void contained in a large one can still pass this
selection criterion, and we expect the fraction of such sub-voids to increase for smaller voids.

We choose one of the major axes of the simulation box as the line-of-sight direction and perturb 
halo positions with their peculiar velocities assuming the plane-parallel approximation: 
\begin{equation}
s=r+v_{z}/aH,
\end{equation}
where $s$ and $r$ is the redshift-space and real-space comoving coordinates of haloes. $v_z$ is the line-of-sight peculiar velocity. 
We view our voids along the three major axes of the simulation box, which effectively increases our sample by a factor of 3. 
All figures in our paper are made in this way unless specified. 

With the halo field in redshift space, we measure the redshift-space
void-halo correlation function $\xi^s{(\sigma, \pi)}$. We rescale the
correlation function for each void by the void radius along both the
$\sigma$ and $\pi$ directions. Examples of the stacked correlation
function are shown in the top-left panel of Fig.~\ref{Fig:LinearModel1}, 
\ref{Fig:LinearModel2}, \ref{Fig:LinearModel3} \& \ref{Fig:measurement}. 
For all cases, there are very good agreements between the black contours from simulation with the 
white contours from the linear model. A more quantitative 
comparison of model and numerical data will be conducted in Section~\ref{Sec:Measurement}. 

\section{Distortion patterns: Elongation vs. Flattening}
Before we move on to write down estimators for the growth based on the above models, 
we first use these models and the simulation to understand the distortion pattern of voids in redshift space, i.e. whether a spherical void 
should appear elongated or flattened along the LOS. This is particularly relevant 
for the AP test, which  
relies on measuring the apparent 
size of voids along the LOS and transverse to it.  

In the linear model, the redshift-space void configuration depends on the real-space void profile, the radial velocity profile as well as its gradients [Eq.~(\ref{Eq:2}) \& (\ref{Eq:4})]. 
From the linear model [Eq.~(\ref{Eq:zspaceProfile})], the distortion pattern of voids depends solely on the real-space void profiles, as the real velocity and its gradients are both determined by the void profile. It can be understood by examining the correlation function in the two extreme directions: LOS versus transverse to the LOS. 
When $\mu^2 \ll 1$, the correlation function can be approximated by its transverse version, i.e.
\begin{equation}
\label{Eq:Tra}
\xi_{\rm vm}^s ({\bf r})=\delta(r)+\frac{1}{3}f\bar{\delta}(r);
\end{equation}
When $\mu^2 \approx 1$, the correlation function can be approximated by its LOS version, i.e. 
\begin{equation}
\label{Eq:LOS}
\xi_{\rm vm}^s ({\bf r})=(1+f)\delta(r)-\frac{2}{3}f\bar{\delta}(r).
\end{equation}
Since $\bar{\delta}(r)$ is always negative within the void radius by definition, the second term in Eq.~(\ref{Eq:Tra}) is of the same sign as the first term, and it acts to deepen the density contrast in the transverse direction. The second term in Eq.~(\ref{Eq:LOS}), including the negative sign has the opposite sign with respect to the first term, which itself is boosted by a factor of $(1+f)$. Hence, the amplitude of clustering along the LOS can be enhanced or reduced, depending on the competing amplitudes of $f\delta$ versus $\frac{2}{3}f\bar\delta$. For haloes, $f\delta$ becomes $\beta\delta$, 
and for massive haloes with a linear bias of $\approx 2$, $\beta$ is relatively small. The amplitude of $f\delta$ is usually smaller than $\frac{2}{3} \bar{\delta}$ at $r<r_{\rm v}$. Hence the second term of Eq.~(\ref{Eq:LOS}) cancels part of the first term, making the density contrast shallower along the LOS. Both these two effects act to flatten the void. Therefore, voids usually appear flattened in redshift space within the void radius. This is visualised in the top-right panels of Figs~\ref{Fig:LinearModel1}, \ref{Fig:LinearModel2} \& \ref{Fig:LinearModel3}. 
At $r<r_{\rm v}$, all the three stacked voids appear flattened, though the amplitudes of the distortion patterns are relatively weak and difficult to 
see on the top-left panels where the full correlation functions from simulation and the best-fit models are shown. 

The distortion feature can also be understood by differencing the transverse correlation function by its LOS version, i.e. subtracting Eq.~(\ref{Eq:Tra}) by Eq.~(\ref{Eq:LOS}), which yields 
$f[\bar\delta(r)-\delta(r)]$. Within $r_{\rm v}$, $\delta(r)$ and $\bar \delta(r)$ are both negative and the amplitude of the latter is larger. 
Therefore, $f[\bar\delta(r)-\delta(r)]<0$, i.e. the correlation function is more negative at the transverse direction than along the LOS. So the void is flattened. 

At radii greater than $r_{\rm v}$, the distortion pattern depends on the
type of voids.  Uncompensated voids are typically embedded in
larger-scale underdense environments, which also expand in an unbound manner. These are
the so-called voids-in-voids
\citep{ShethVDWeygaert2004}. Here, $\bar{\delta}(r)$ is negative up to large
radii. If the differential profile $\delta(r)$ also stays negative at
large radii, the same flattening feature is expected. But if $\delta(r)$
turns positive at some large radius, the distortion pattern becomes
elongated. For uncompensated voids, this happens typically at large
radii. So the general picture for uncompensated voids in redshift
space is flattening at relatively small radii (greater than $r_{\rm v}$)
and elongation at the very large radii when $\delta(r)$ becomes
positive. Fig.~\ref{Fig:LinearModel3} gives an example of this
kind. On the bottom-left panel, the cumulative void profile
$\bar{\delta}(r)$ is negative up to $4r_{\rm v}$. The differential void
profile indicated by the red dots and red dashed curves remains
negative at $r<2r_{\rm v}$ and it become weakly positive at $r>2r_{\rm v}$. In the
top panel, the flattening feature is seen up to twice the void
radius; it then becomes elongated at larger radii as expected.

In contrast, over-compensated voids are likely to be embedded in overdense 
environments: these are voids-in-clouds
\citep{ShethVDWeygaert2004}. Here, $\bar{\delta}(r)$ is negative within
the compensation radius $r_{\rm com}$,
becoming positive for $r>r_{\rm com}$.  The differential profile
$\delta(r)$ must then become positive at $r<r_{\rm com}$. Therefore,
at $r<r_{\rm v}$, the void appears flattened; between $r_{\rm v}<r<r_{\rm com}$ it
is elongated. At $r>r_{\rm com}$, the shape becomes
flattened again as both $\delta(r)$ and $\bar{\delta}(r)$ remain
positive, which is essentially the same as the flattening feature in overdensities. An example is given in Fig.~\ref{Fig:LinearModel1}, where we
have selected voids from our simulation with $\bar{\delta}(r)>0$ at
$r=r_{\rm v}$ to make sure they are voids surrounded by large-scale overdense 
environments. The over-compensation property of the stacked
voids is also reflected by the fact that the mean radial velocity
becomes negative at $r>1.7r_{\rm v}$, as shown in the bottom-right panel of
Fig.~\ref{Fig:LinearModel1}.

When adding a dispersion in velocity according to the quasi-linear
model shown in Section~\ref{Quasi}, we find that the correlation
functions are weakly elongated along the LOS, and the effect is
stronger at $r<r_{\rm v}$. In some cases where the dispersion is dominant
over the mean streaming velocity, the interior of voids becomes
elongated rather than flattened. At larger radii, the impact of
velocity dispersion is found to be negligible. It is worth noting that, according to the quasi-linear model, 
the interior of the void may also appear elongated when the amplitude of $\delta$ is relatively large, i.e. $\delta \sim -1$, where the linear model is no longer accurate.

In summary, peculiar velocities cause complex distortion patterns for
voids in redshift space, which can largely be understood using linear theory.
Voids usually appear flattened within the radius of underdensity
according to linear theory, but can be elongated when velocity
dispersion is non-negligible; they may be elongated or flattened at
large radii. This behaviour is mostly determined by the density profile, as the impact
of velocity dispersion is very small at larger radii. In some cases,
the transition between flattening and elongation is very abrupt. 
This is different from the relatively simple distortion
    patterns for overdensities, where infall streaming motion causes
    flattening on large scales
and random velocity dispersion induces Finger-of-God-like elongation on small scales.
This suggests that the effect of redshift-space distortion on the AP
measurement using voids depends strongly on both the void population
and the radius where the size of the void is taken. It is perhaps
over-simplistic to apply a single stretching factor to correct for
this effect, as was done in \citet{Sutter2014}, given the wide range of
void profiles encountered in practice.

The fact that the impact of velocity dispersion is small at large
radii suggest that the linear model without velocity dispersion may
be sufficient to provide good fit for the correlation function at
$r>r_{\rm v}$.
As we will show in the next section,
allowance for velocity dispersion is needed only when fitting at $r<r_{\rm v}$,

\section{measuring the growth rate}
\label{Sec:Measurement}
\subsection{The growth estimator}
\label{Sec:estimator}
Following \cite{Hamilton1992,Hamilton1998}, we can write down the following pair of equations for the 
void-mass correlation function:
\begin{align}
\xi_{0}^s(r)-\frac{3}{r^3}\int_0^r \xi^s_0(r')r'^2dr' &=\left(1+\frac{f}{3}\right)[\xi(r)-\bar \xi(r)] \\
\xi_{2}^s(r) &=\frac{2f}{3} [\xi(r)-\bar \xi(r)] ,
\end{align}
where $\bar\xi(r)=(3/r^3)\int_0^r \xi(r')r'^2dr'$. Since the hexadecapole is zero, we have 
\begin{align}
\label{Eq:zspaceCF}
\xi^s(r, \mu)&= \xi_0^s(r)+\frac{3\mu^2-1}{2}\xi^s_2(r) \\ \nonumber
&= \left(1+\frac{f}{3}\right)\xi(r)+\frac{f(3\mu^2-1)}{3}  [\xi(r)-\bar \xi(r)] \\  \nonumber
&=\xi(r)+\frac{1}{3}f\bar \xi(r)+f\mu^2[\xi(r)-\bar\xi(r)]
\end{align}
The above expression matches to Eq.~(\ref{Eq:zspaceProfile}) because the void-mass correlation function $\xi(r)$ 
is essentially the real-space density profile of voids $\delta(r)$; thus $\bar \xi(r)=\bar{\delta}(r)$.

With the above pair of moments of the correlation function, we can estimate the growth rate using the estimator
\begin{align}
\label{Eq:Estimator}
\tilde G(f) &= \frac{\xi_{2}^s(r)}{\xi_{0}^s(r)-\frac{3}{r^3}\int_0^r \xi^s_0(r')r'^2dr'} \\
&= \frac{2f}{3+f}.
\end{align}
The multipoles of correlation function can be obtained by
\begin{equation}
\xi^s_{\ell}(r)=\int_0^1\xi^s(r, \mu)(1+2\ell)P_{\ell}(\mu)d\mu,
\end{equation}
where $P_{0}(\mu)=1$ and $P_{2}(\mu)=\frac{1}{2}(3\mu^2-1)$.

\begin{figure*}
\begin{center}
\scalebox{0.5}{
\includegraphics[angle=0]{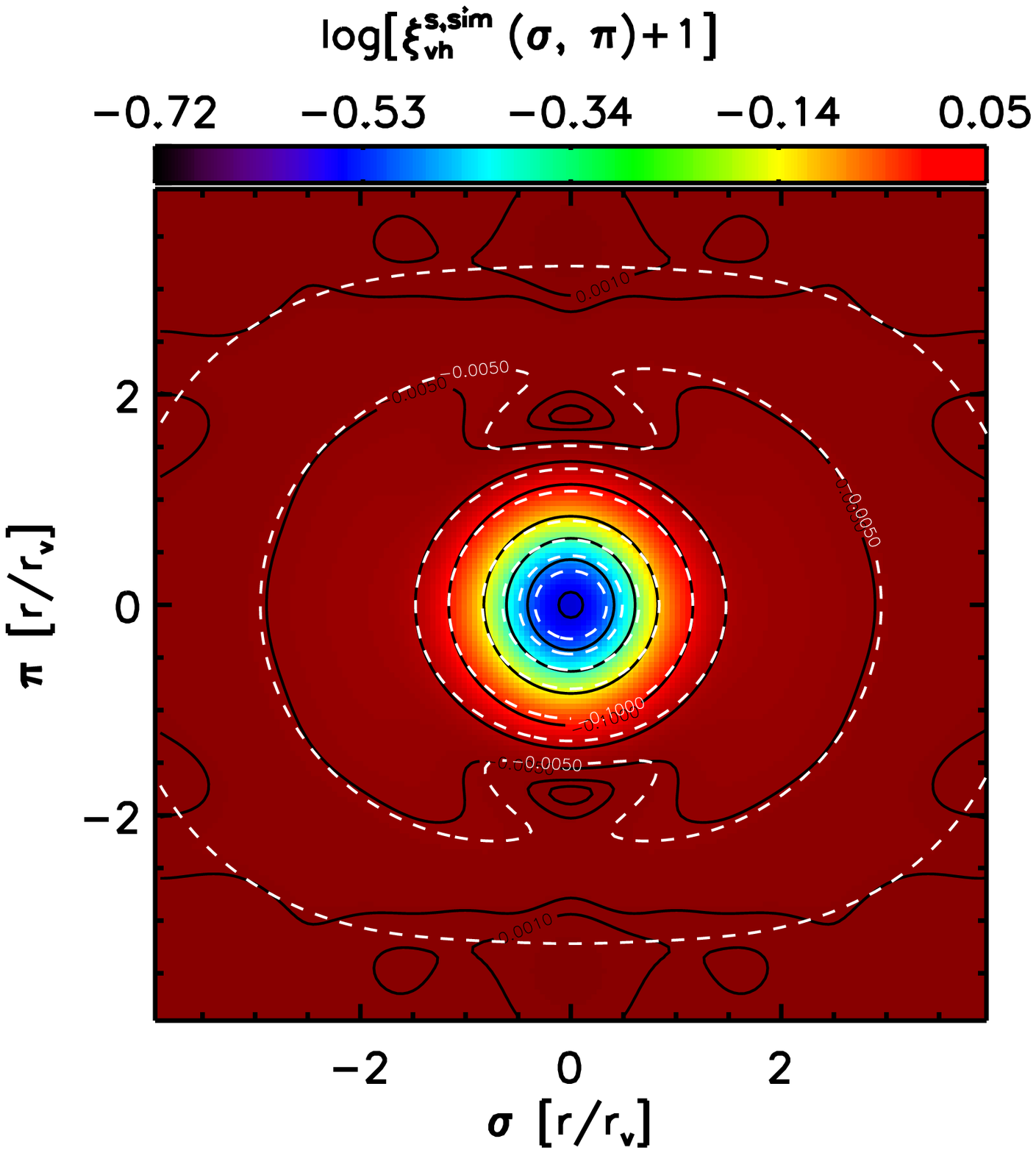}
\hspace{-2.0cm}
\includegraphics[angle=0]{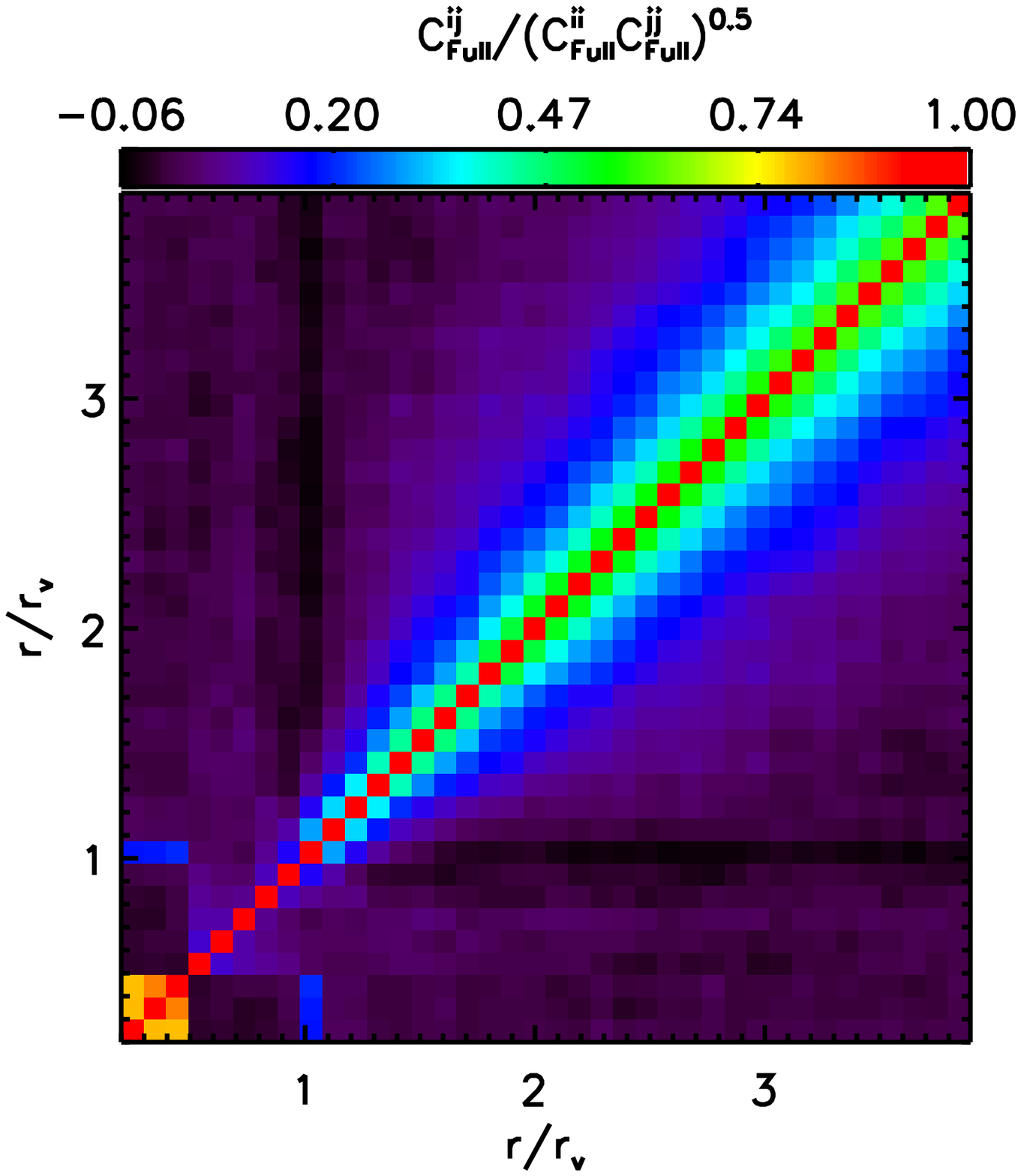}}
\scalebox{0.42}{
\hspace{-1 cm}
\includegraphics[angle=0]{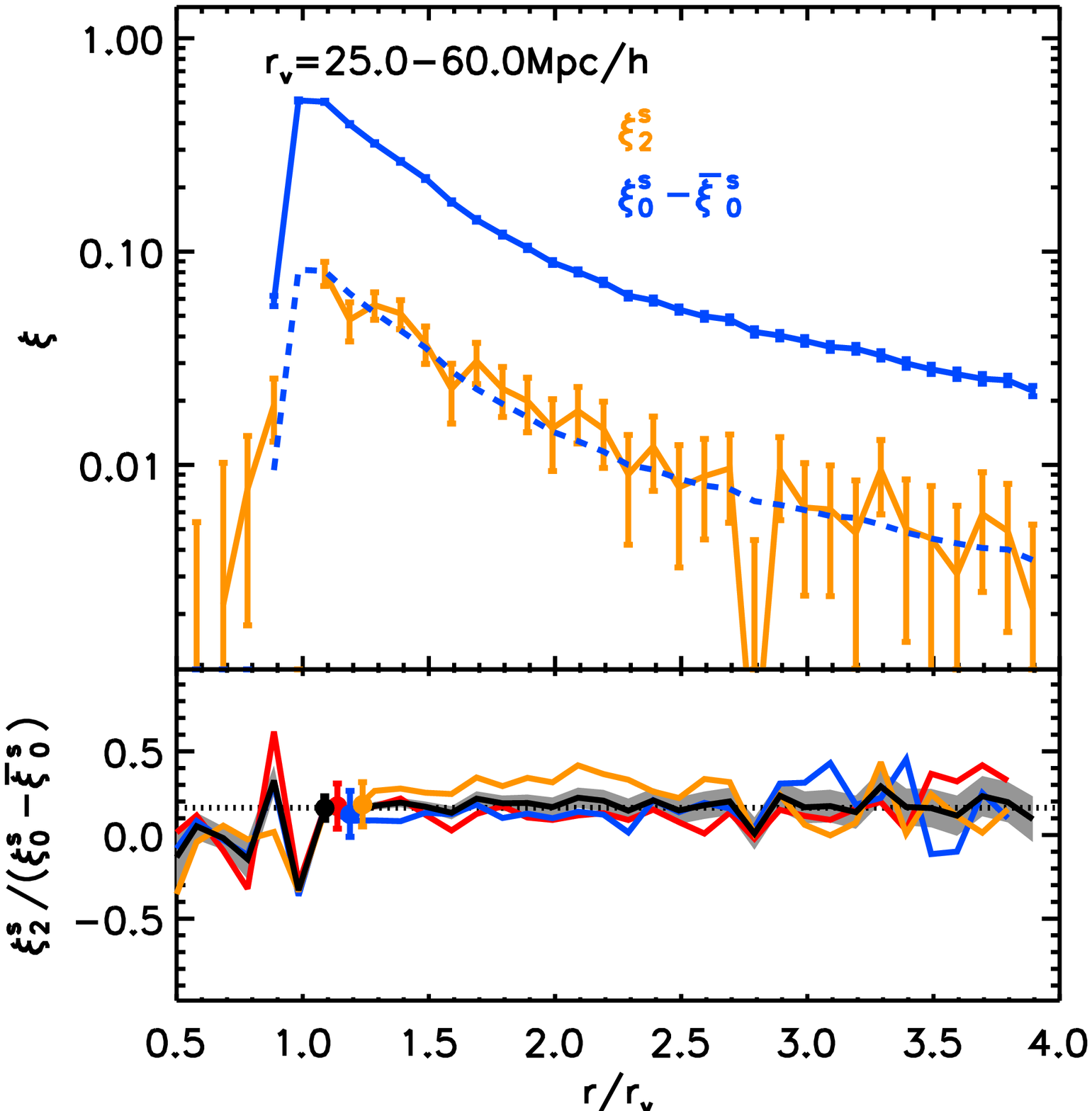}
\hspace{2cm}
\includegraphics[angle=0]{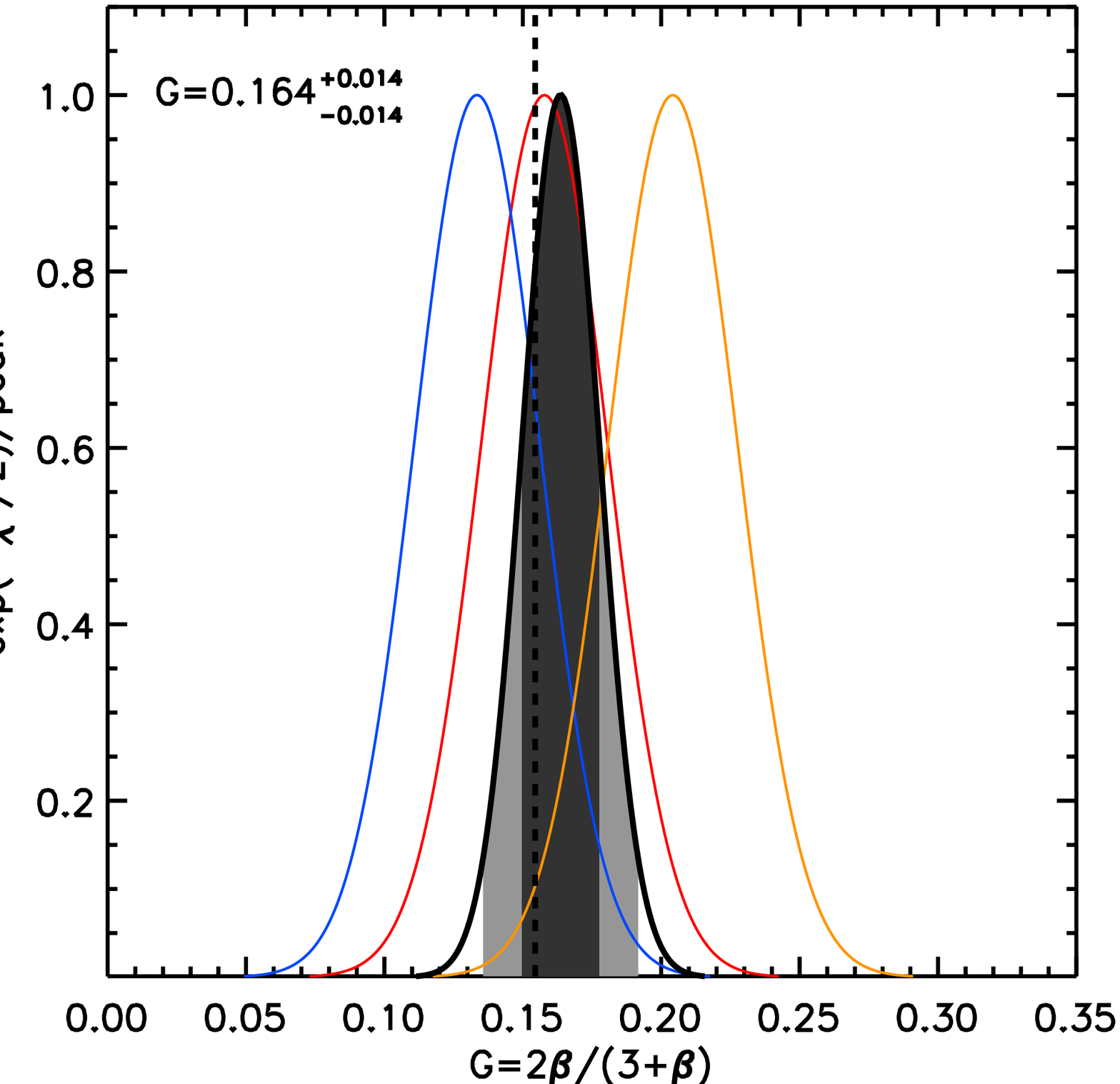}
}
\caption{Top-left: The void-halo correlation function in redshift
  space. The black contours give results from the simulation and the white
  contours are the best-fit linear model. Top-right: Cross-correlation
  coefficients of the covariance matrix shown in
  Eq.~(\ref{Eq:Cov_Full}). Bottom-left: the upper panel shows the monopole
  and quadrupole moments of the redshift space correlation
  function. The lower panel shows the ratio of the quadrupole versus
  monopole. The black curve is from taking all measurements of voids
  along three different major axes of the simulation box, with 
  the shaded region showing the error on the mean. The other
  three curves represent results from viewing the
  simulation along three different major axes.  The black filled circle
  with error bars is the best-fit value from viewing voids along three
  different directions.  The red, blue and orange filled circles and errors
  are from individual viewing directions. They are slightly offset from
  each other to aid visibility. The error bars have been amplified by a
  factor of 5 to be visible. Bottom-right: the normalised likelihood
  for the best-fit $G$ values. The dark grey regions are the 1 and
  2-$\sigma$ ranges for results from measuring voids in redshift space
  along three different major axes of the simulation box. The other
  three colours curves are likelihoods from viewing voids along three
  different directions.}
\label{Fig:measurement}
\end{center}
\end{figure*}

\begin{figure*}
\begin{center}
\scalebox{0.5}{
\includegraphics[angle=0]{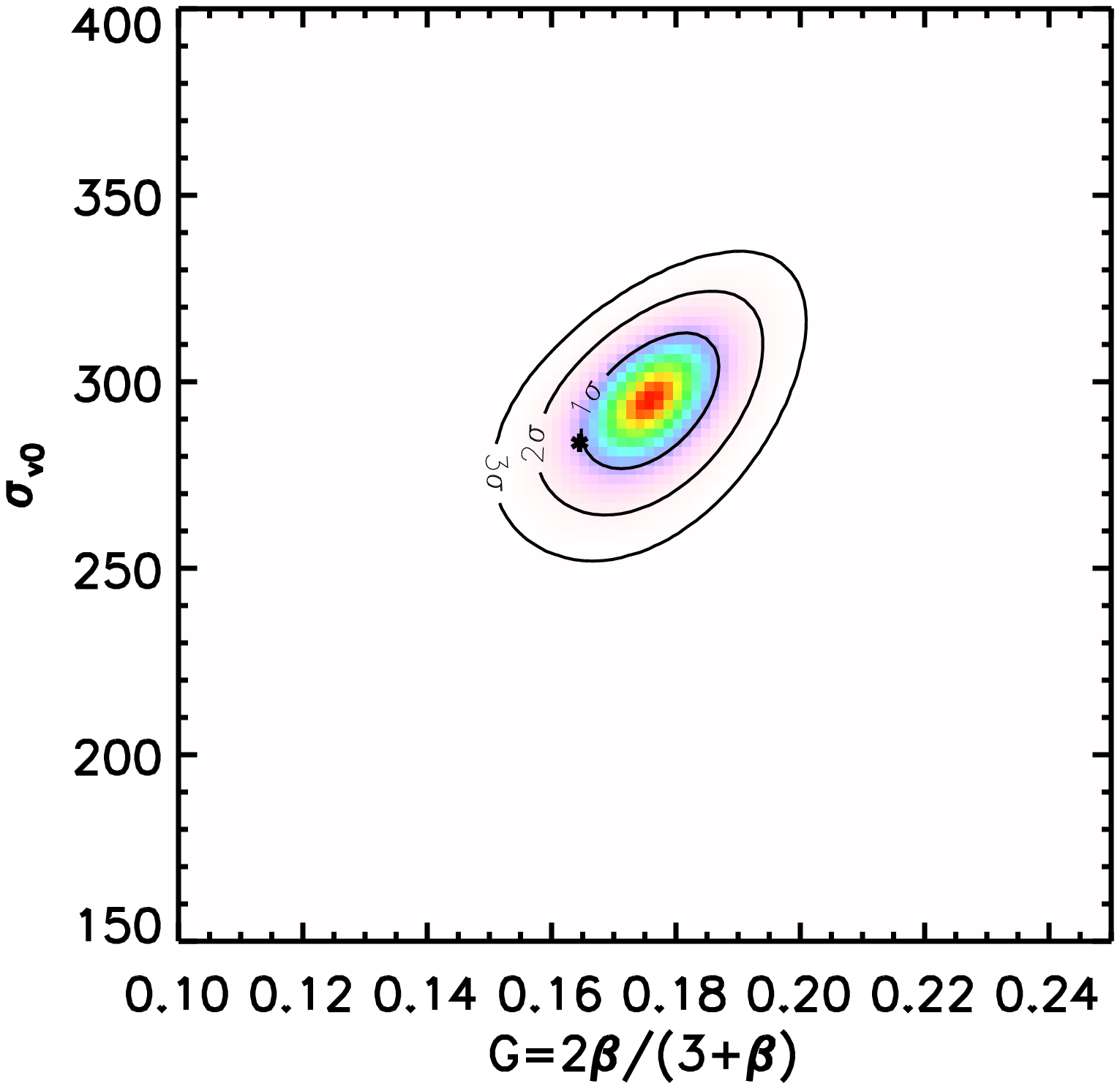}
\hspace{-2.0cm}
\includegraphics[angle=0]{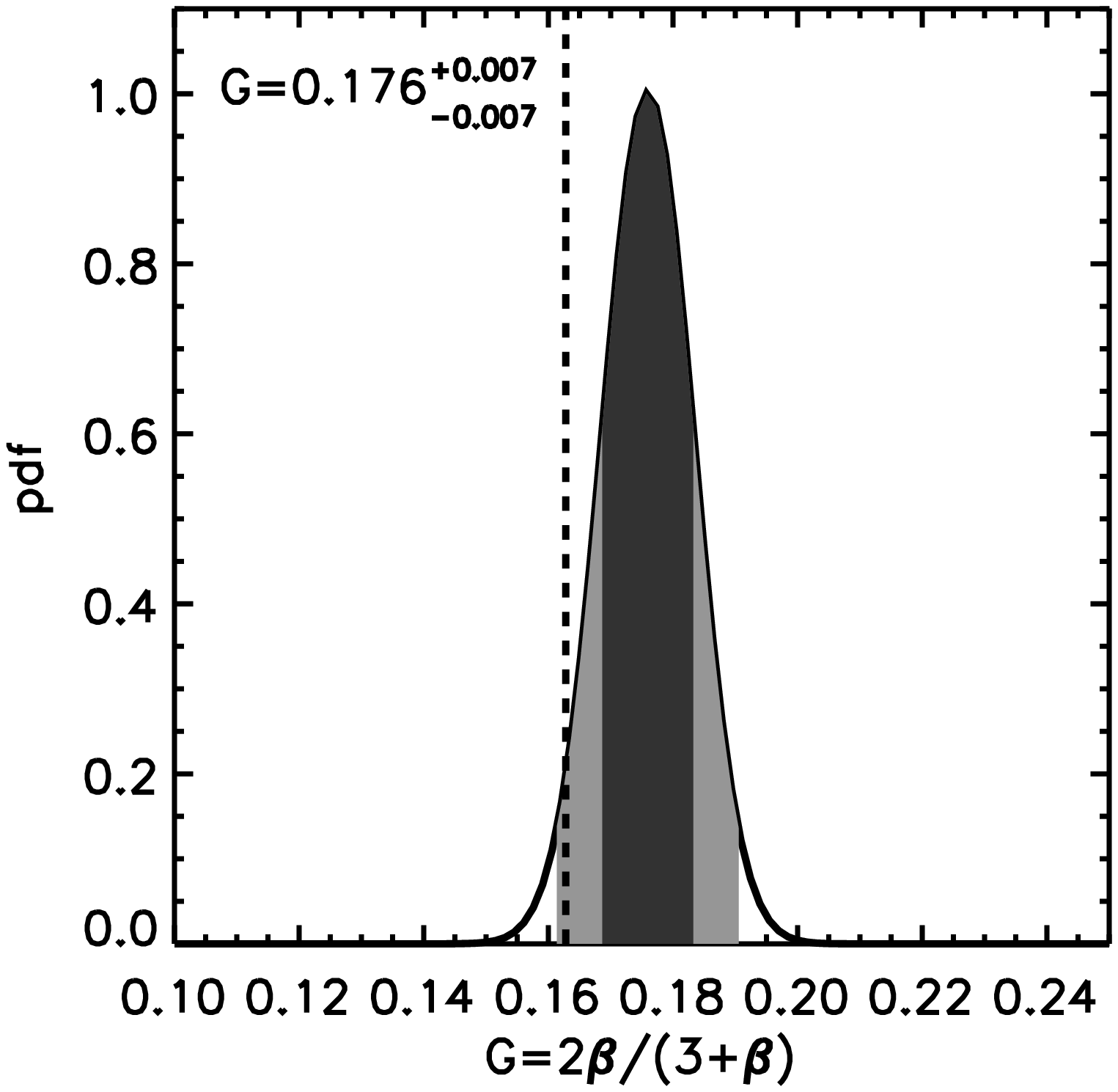}}
\scalebox{0.42}{
\hspace{-1 cm}
\includegraphics[angle=0]{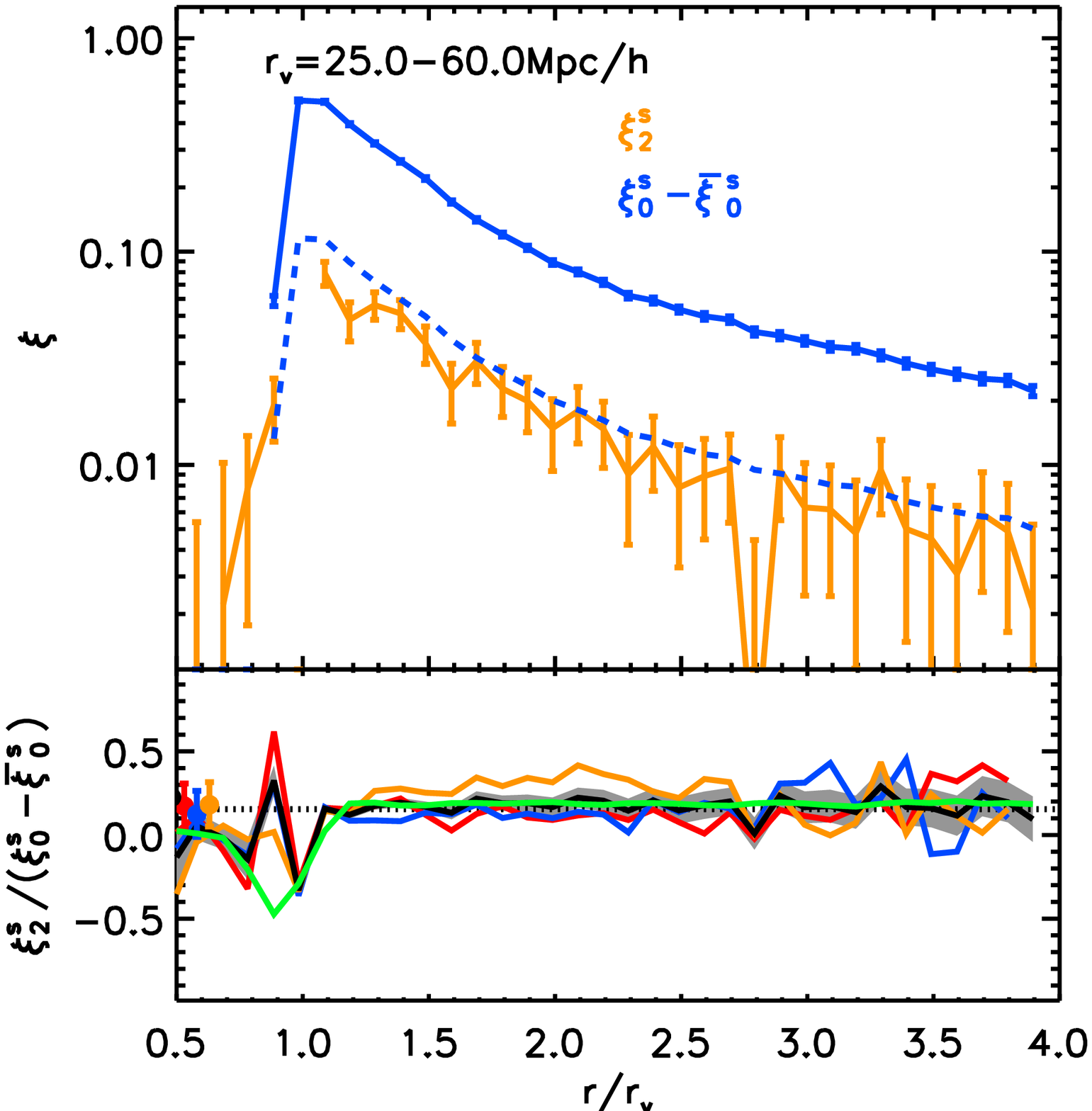}
\hspace{2cm}
}
\caption{Top-left: $\sigma$ contours for the two free parameters $[\sigma_0, G]$, where $\sigma_0$ is the amplitudes of the velocity dispersion profile taken to follow an error
function form, i.e. $\sigma_{\rm v}(r)=\sigma_0 {\rm erf}(r/r_{\rm v})$. The star indicates fiducial values. Top-right: constraint for the growth parameter $G$ after marginalising over $\sigma_0$. 
Bottom: similar to the bottom-left panel of Fig.~\ref{Fig:measurement} but with the best-fit model curve including velocity dispersion shown in green.}
\label{Fig:measurement_Dispersion}
\end{center}
\end{figure*}

We can also express these results in Fourier space, again assuming the
plane-parallel approximation. Following \cite{Hamilton1998}, the
redshift-space overdensity is mapped into the real-space overdensity
via the redshift-space operator $\bf{S}$, $\delta^s({\bf
  r})={\bf S} \delta({\bf r})$, where in the linear regime, ${\bf
  S}=1+f (\partial^2/\partial r_c^2)\bigtriangledown^{-2}$ and in
Fourier space ${\bf S}=1+f\mu_{\bf k}^2$ with $(\partial^2/\partial
r_c^2)\bigtriangledown^{-2}=k^2_z/k^2=\mu_{\bf k}^2$ and $\mu_{\bf
  k}=\hat{\bf z} \cdot \hat{\bf k}$.  Since the bulk motion of the
void will not show up in the void-galaxy correlation, and we only
require the peculiar motion of galaxies with respect to the void
centre, we only need to apply one redshift distortion operator to the
real-space density contrast to obtain the void-mass cross-power
spectrum in redshift space:
\begin{equation}
P_{\rm vm}^s({\bf k})=(1+f\mu_{\bf k}^2)P_{\rm vm}(k).
\end{equation}
This Fourier expression may seem inconsistent with previous work on
cross-correlations in redshift space, e.g. \cite{Mo1993}. Normally,
one would argue that the linear Fourier-space density fluctuation for a single population is
$\tilde\delta = b\tilde\delta_m(1+\beta\mu^2)$, so that the cross-power for
two differently biased tracers is
$P_{12}=b_1b_2P_m(1+\beta_1\mu^2)(1+\beta_2\mu^2)$. Thus apparently a
hexadecapole is expected unless one of the bias values diverges;
this is certainly not the case for voids from our simulation, where the large-scale linear
bias is approximately $-2$, as we have found from measuring the real space void bias from $b_{\rm v}=\xi_{\rm vm}/\xi_{\rm mm}$. 

This apparent discrepancy is resolved as follows. First, note that the usual linear
expression assumes a local bias, with the density of the tracer proportional to matter density;
this will not apply on scales of the size of the objects concerned, which is exactly where
we are working. This point can be made in more detail using the language of the halo
model, where we would write the void-galaxy cross-correlation in real space as
\begin{equation}
\xi_{\rm vg}(r)= \Delta_{\rm v}(r) + b_{\rm v} b_g \xi_{\rm mm}(r),
\end{equation}
where the 1-void term $\Delta_{\rm v}(r)$ is the mean density contrast
around the voids.  When casting this into redshift space, the
2-void term $b_{\rm v} b_{\rm g} \xi_{\rm mm}(r)$  indeed generates a
hexadecapole as above. As for the 1-void term, normally 
in the halo model one would
consider this to be predominantly distorted by virialized random
velocities.
But for voids, the dominant internal velocities are coherent outflow, and
these generate a quadrupole in the cross-correlation function.
This is true even
though we have adopted linear theory to calculate the outflow
velocity, whereas `linear' would seem to imply the 2-void term.

In the usual application of this equation to haloes, the 1-halo term
dominates at small separations because the central overdensity of haloes is
high, whereas $\xi_{\rm mm}$ is of order unity there because it is the linear
correlation function. This argument is less applicable here, since
$\Delta_{\rm v}$ cannot be below $-1$; but nevertheless the 2-void term
is unimportant near $r=r_{\rm v}$.
The reason for this
is {\it exclusion\/}, which is a condition that applies also to
haloes: for objects of a finite size, we cannot expect to find correlated
pairs where the centres lie very close together. Thus the
above halo-model expression is inaccurate, and the hexadecapole-generating
2-void term will not contribute on the scales around $r_{\rm v}$ where
we concentrate our analysis.
This can be seen in the good
  agreement between our linear model with simulation
  (shown in Figs~\ref{Fig:LinearModel1}, \ref{Fig:LinearModel2} \&
  \ref{Fig:LinearModel3}), verifying empirically that the signal
is dominated by the pure-quadrupole 1-void term 
\footnote{We have also found that the `linear' void bias $b_{\rm v}$ increases with decreasing 
void radius, consistent with what was found in \cite{Chan2014} and \cite{Hamaus2014}. 
In principle, at a certain void radius where the void population has $b_{\rm v}\approx0$, the amplitude $\beta_2$ from the void-void contribution will be large, so
the hexadecapole is also expected to be large. However, the (quasi-) linear model may also break down for small void radii, where the underlying 
density field as well as the galaxy or halo biases are expected to be non-linear. We leave this as an open question to be investigated in future work.}.

We can now proceed to decompose the redshift-space power spectrum into 
Legendre polynomials, where the only two non-zero moments are $\ell=0$ and 2:
\begin{align}
P_{0}^s(k)&=\left(1+\frac{f}{3}\right)P_{\rm vm}(k) \\
P_{2}^s(k)&=\frac{2f}{3} P_{\rm vm}(k).
\end{align}
The Fourier-space estimator is therefore:
\begin{equation}
\tilde G(f) = \frac{P_{2}^s(k)}{P_{0}^s(k)} = \frac{2f}{3+f},
\end{equation}
which gives the same answer as the real-space version. In the following, 
we will use an N-body simulation to test the linear estimator in configuration space.  

\subsection{Testing the estimator with simulation}
We use a sample of voids defined using the halo number-density 
field to test the performance of the estimator in 
configuration space. 
We have $N=4560$ voids with $r_{\rm v}>25 \Mpc$. The stacked void-halo correlation function for voids greater than $25\Mpc$ is shown in the top-left panel of 
Fig.~\ref{Fig:measurement}. The stacked void is underdense within $r_{\rm v}$.  At the edge of the void radius, the
overdensity of haloes rises steeply towards positive values, but it does not 
become positive until $r\approx 4r_{\rm v}$. The stacked void is slightly flattened 
within $r_{\rm v}$ and it remains flattened until $r\approx 4r_{\rm v}$ where $\delta(r)$ becomes positive. 
This is expected from the linear-theory analysis in the previous section, 
which attributed the distortion pattern to the competing amplitudes of the local density, 
radial velocity as well as its gradient. 

We decompose each correlation function into monopole and
quadrupole components. In each radial bin, we extract the correlation function as a
function of $\mu^2$.  The correlation function becomes noisy in the
vicinity of the $\mu=1$ LOS direction as the volume per
pixel becomes smaller. We remove the part of the correlation function
within 0.5$r_{\rm v}$ around the middle along the LOS in order to reduce the
noise, and fit $\xi(r,\mu)$ at each $r$ with Eq.~(\ref{Eq:zspaceCF}). We then
integrate the best-fit monopole $\xi_0^s$ using Eq.~(\ref{Eq:xibar}) to obtain the
monopole term as shown in the denominator of Eq.~(\ref{Eq:Estimator}).

The monopole-quadrupole
decomposition is conducted for all the individual voids. In
principle, we can obtain $4560\times 3$ estimates of the growth factor $\tilde
G$ by taking ratios between quadrupoles and monopoles. In practice,
taking ratios of noisy data is not optimal. To avoid this, we
effectively treat the measured monopole as our model, and rescale it
by varying a constant $\tilde G'$ until it is best matched by the
measured quadrupole, i.e.
\begin{equation}
\xi_{2}^s(r)  = [\xi_{0}^s(r) - \bar{\xi_{0}^s}(r)] \tilde G'.
\end{equation}
This can be treated as a minimum-$\chi^2$ fitting procedure. Note that in doing this we made no further assumption beyond linearity, i.e. the quadrupole-monopole ratio must be a constant if linear theory applies. This simple assumption allows us to fit for the growth using the quadrupole and monopole without actually taking their ratio; and, more importantly, without having any prior knowledge of the shape of the void-halo correlation function. Of course, we need to account for the variances and covariances of the monopole and the quadrupole. It is straightforward to show that the full covariance is 
\begin{equation}
{C^{ij}}_{\rm Full}={C^{ij}_2}+\tilde G'_2{C^{ij}_0}-\tilde G' { C^{ij}_{02}}-\tilde G' {C^{ij}_{20}}, 
\label{Eq:Cov_Full}
\end{equation}
where ${\bf C_0}$ and ${\bf C_2}$ are the covariance of the monopole and quadrupole; ${\bf C_{02}}$ and ${\bf C_{02}}$ are the covariance between ${\bf C_0}$ and ${\bf C_2}$; the subscripts $i$ and $j$ represent different radial bins. Each covariance matrix is computed from the $4560\times 3$ individual measurements, 
\begin{equation}
{C^{ij}}=\frac{1}{N^2}\sum_{k=1}^{N}(\xi^i-\bar{\xi^i})(\xi^j-\bar{\xi^j}), 
\end{equation}
where $N=4560\times 3$, $\xi^{i}$ is the monopole or quadrupole at the
$i$-th radial bin and $\bar{\xi}^i$ is the average from the whole sample.
The prefactor of $1/N^2$ in the above expression arises because we want
to consider the errors on the mean $\xi$, not an individual measurement.
$\chi^2$ values are computed using
\begin{equation}
\chi^2=\sum_{ij}^n\Delta^i [{ C^{ij}}_{\rm Full}]^{-1} \Delta^j, 
\end{equation}
where $ {\bf \Delta }$= ${\bm \xi_{2}^s}$  - [$ {\bm \xi_{0}^s}$ -${\bm \bar{{\bm \xi}}_{0}^s}$] $\tilde G'$ is the vector characterising the difference between the quadrupole and the rescaled monopole. 

The top-right panel of Fig.~\ref{Fig:measurement}  shows an example of the full covariance matrix. 
The correlation between neighbouring bins can be at the 50\% level at $r>r_{\rm v}$, dropping to the 
10\% to 20\% level when radial bins are more than 0.2$r_{\rm v}$ apart. This suggests that the overlap of volume among different voids is not so strong in our sample. 

The bottom-left panel of Fig.~\ref{Fig:measurement} shows the
corresponding quadrupole and monopole from averaging over voids viewed
along all the three major axes of the simulation box. While the
amplitude of the quadrupole remains very small at all scales, the
ratio between the quadrupole and monopole is very nearly independent of radius
for $r>r_{\rm v}$.  The measurement inevitably becomes noisier at smaller radii,
as there are fewer haloes there, but in
general the ratio drops in this regime and in some cases
becomes negative. This is expected as the impact of velocity
dispersion is dominant over the linear streaming motion, causing the
correlation function to be less flattened or even to become
elongated. The linear model clearly fails at $r<r_{\rm v}$; in the
following, we perform fitting to the measurement using two different
models at $r>r_{\rm v}$ and $r>0.5r_{\rm v}$ respectively.

For $r>r_{\rm v}$, we adopt the linear model. We obtain a
best-fit value for the growth factor $G$, as indicated by the
  red, blue and orange dots with error bars for a volume of
  1(Gpc/$h$)$^3$ and the black version is for an effective volume of
  3(Gpc/$h$)$^3$. It can be seen that this is unbiased. The
  precision of the measurement of the growth factor $G$ is about 9\%
  for our effective simulation volume of $3(h^{-1}{\rm Gpc})^3$. This translates into a 9\%
  uncertainty in $\beta$.
 
 \begin{figure*}
\begin{center}
\scalebox{0.28}{
\includegraphics[angle=0]{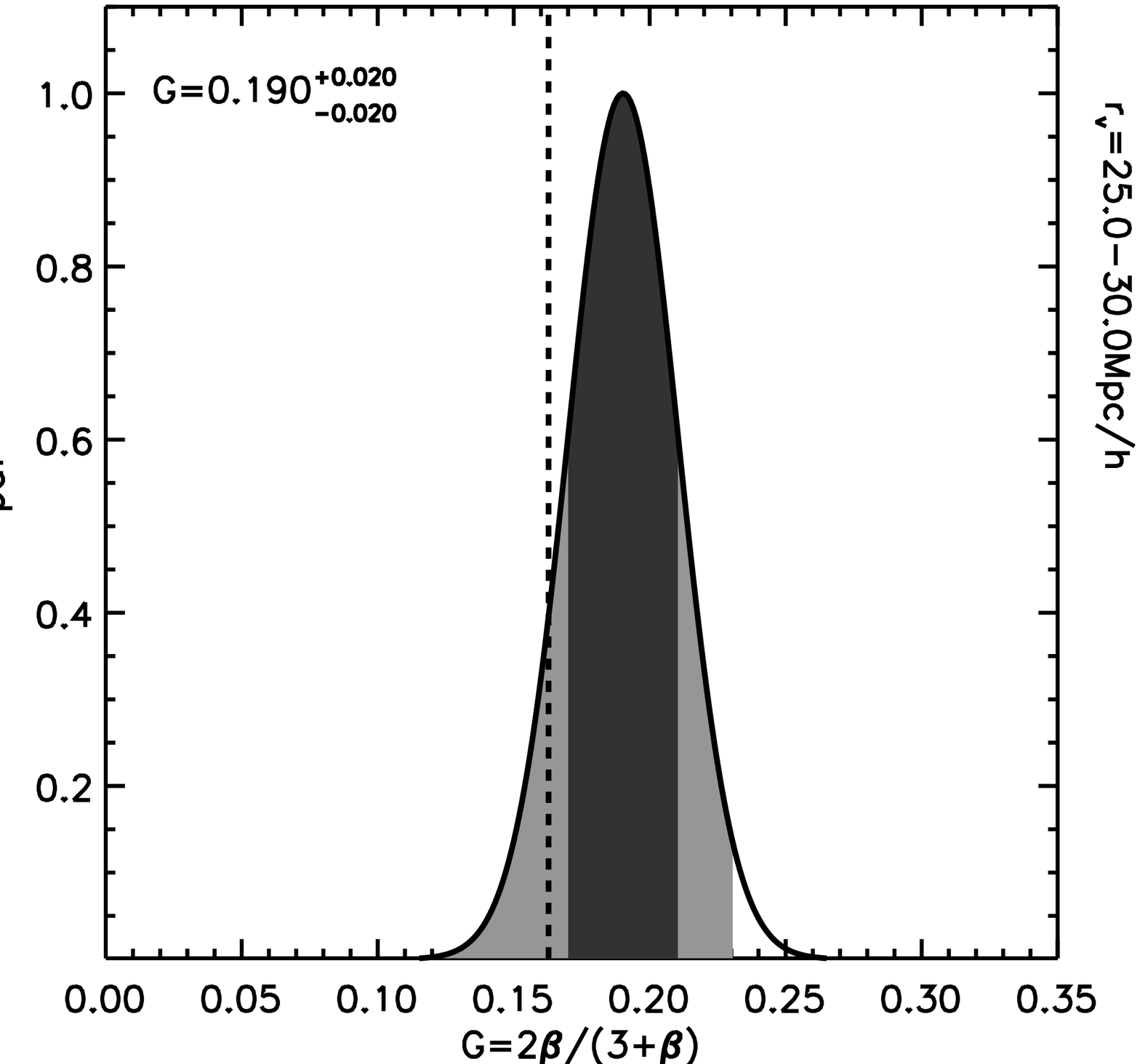}
\hspace{2.5cm}
\includegraphics[angle=0]{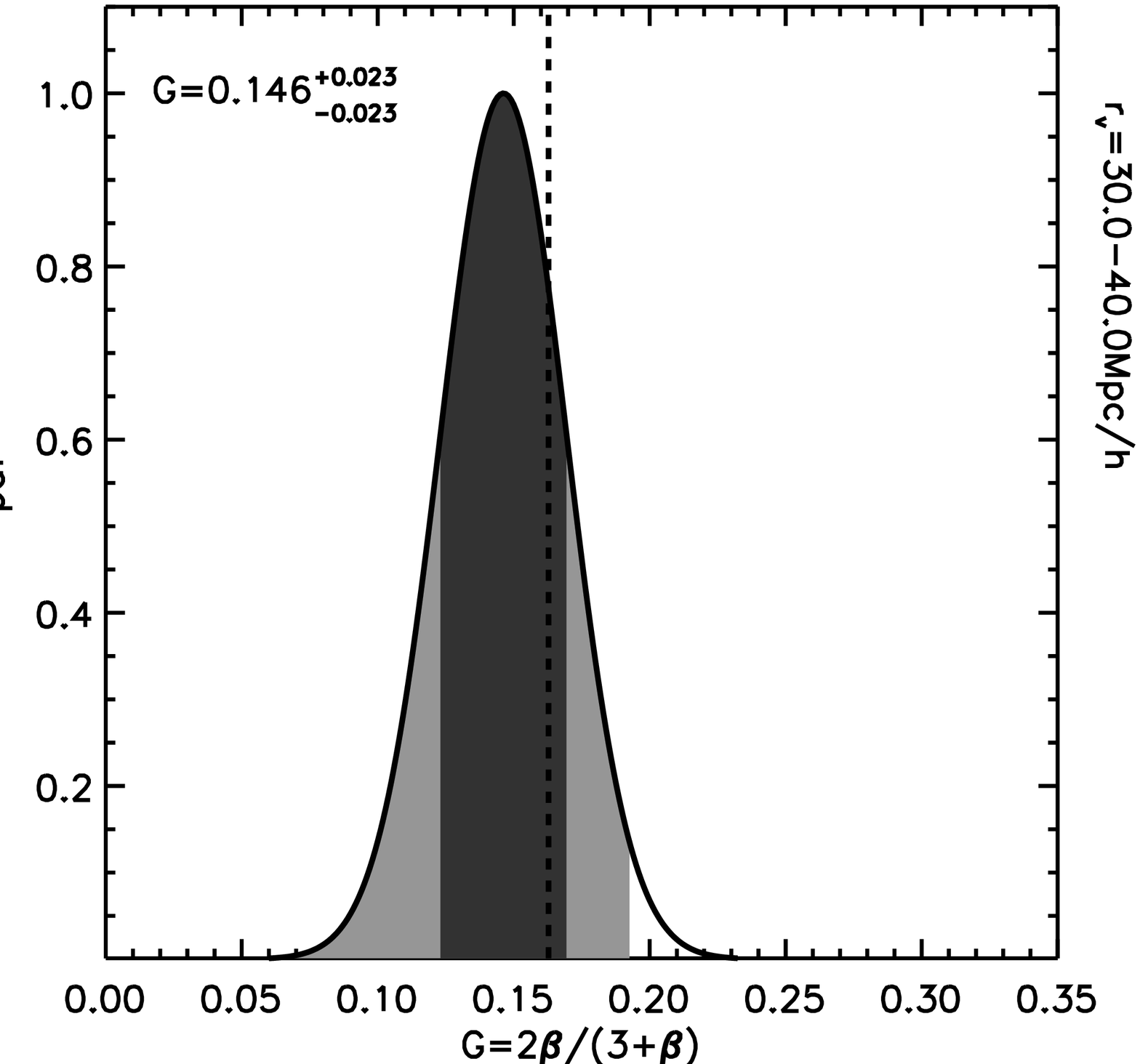}
\hspace{2.5cm}
\includegraphics[angle=0]{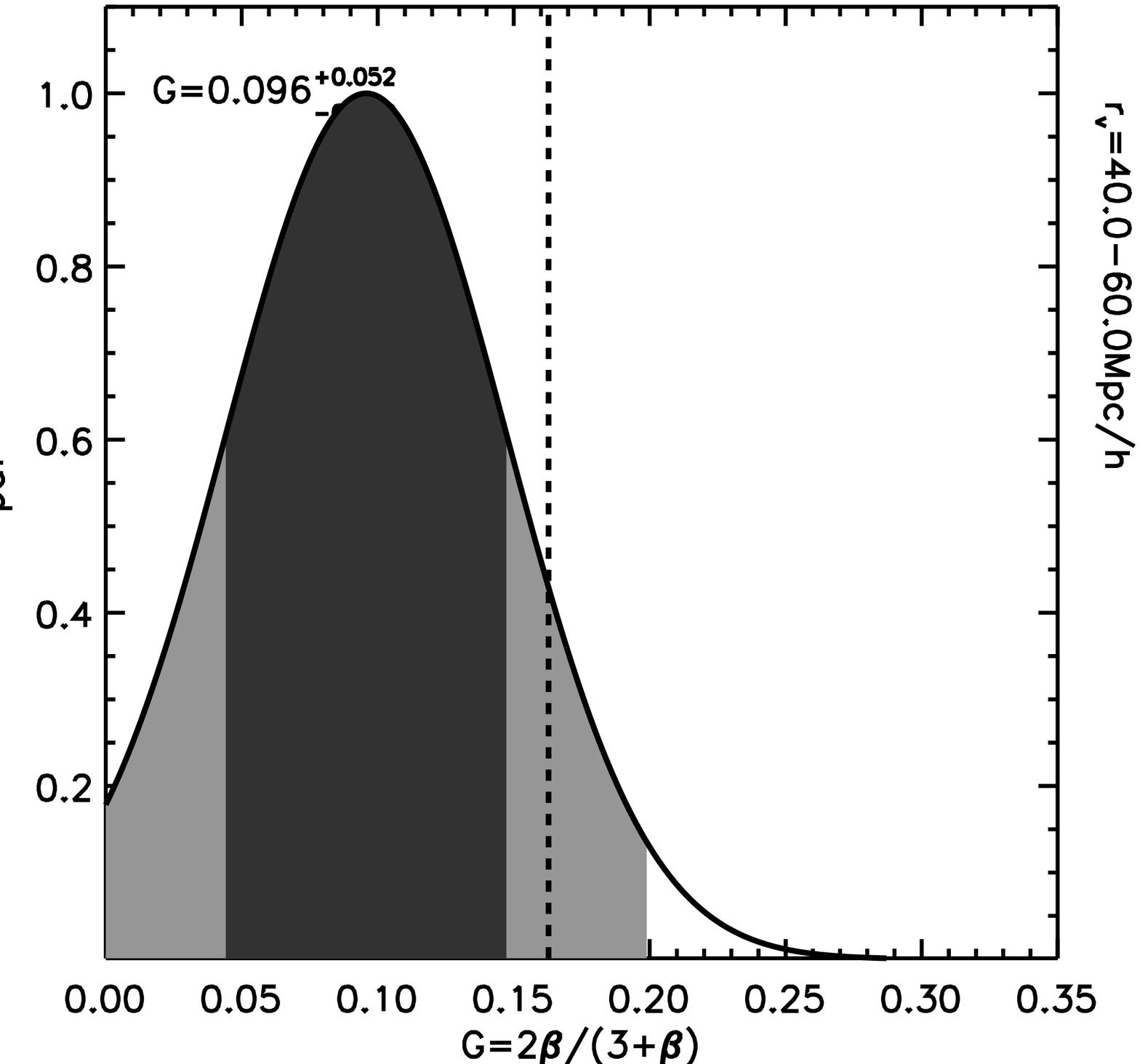}}
\scalebox{0.38}{
\hspace{-1.3cm}
\includegraphics[angle=0]{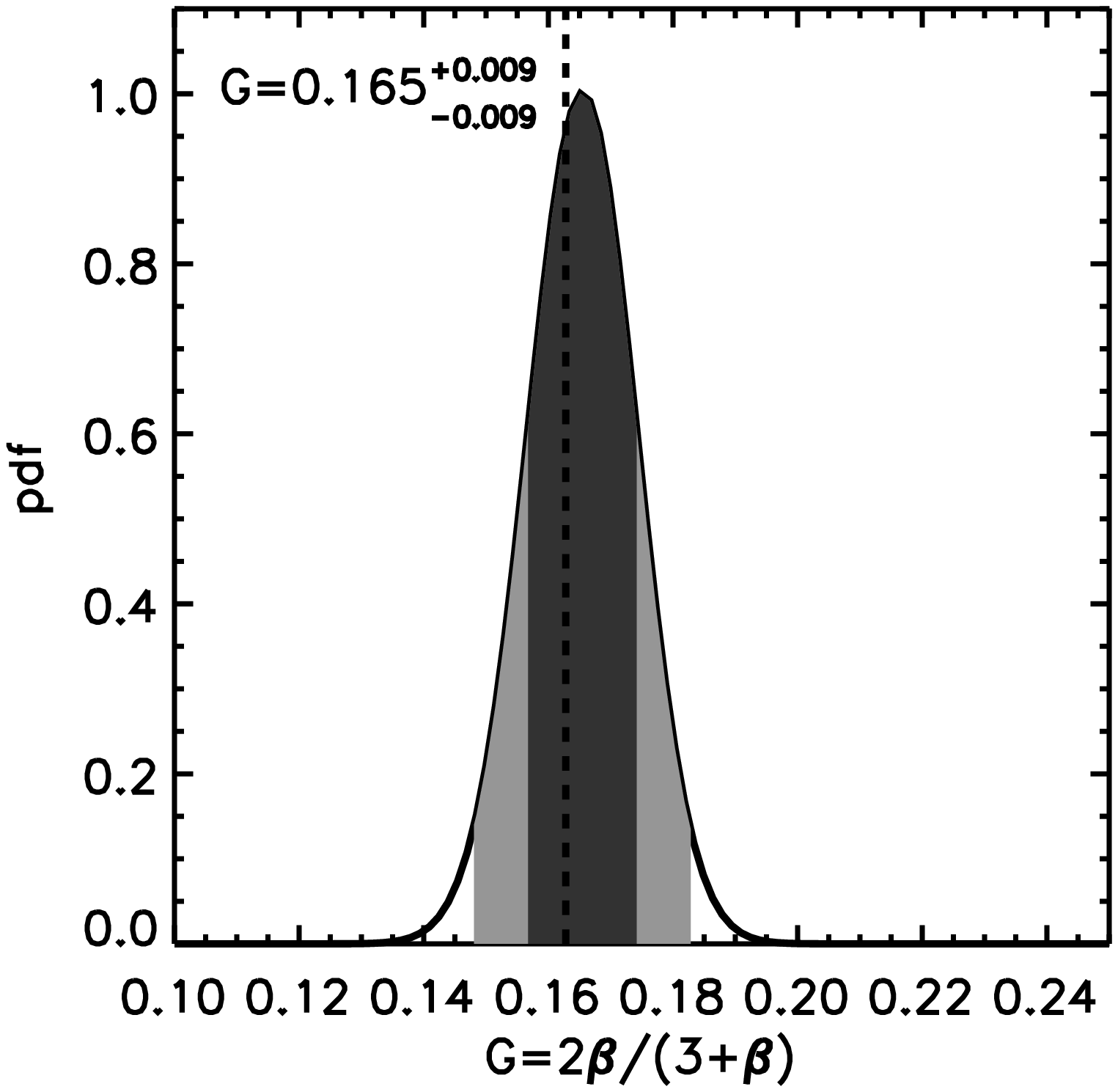}
\hspace{-3.3cm}
\includegraphics[angle=0]{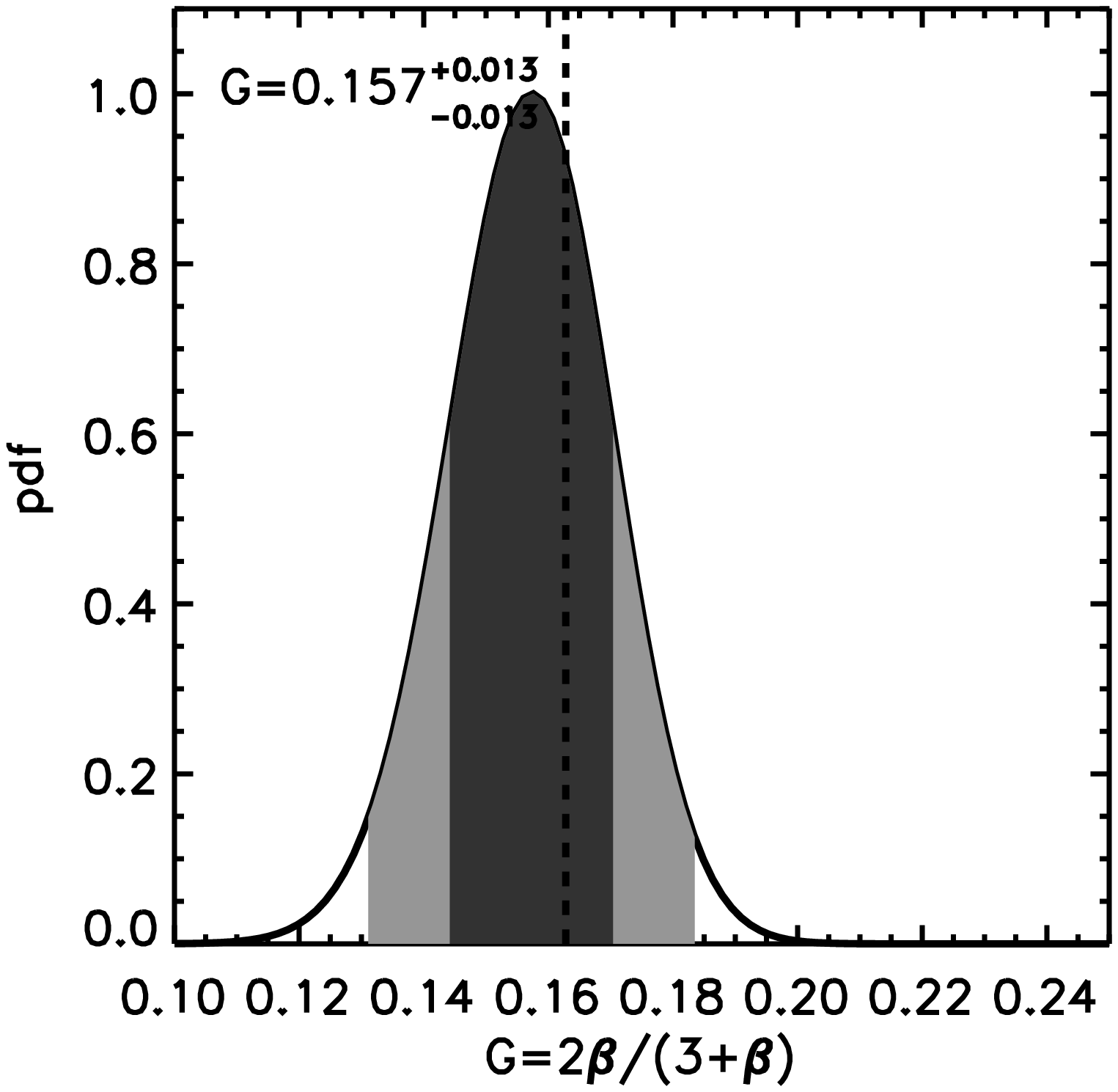}
\hspace{-3.3cm}
\includegraphics[angle=0]{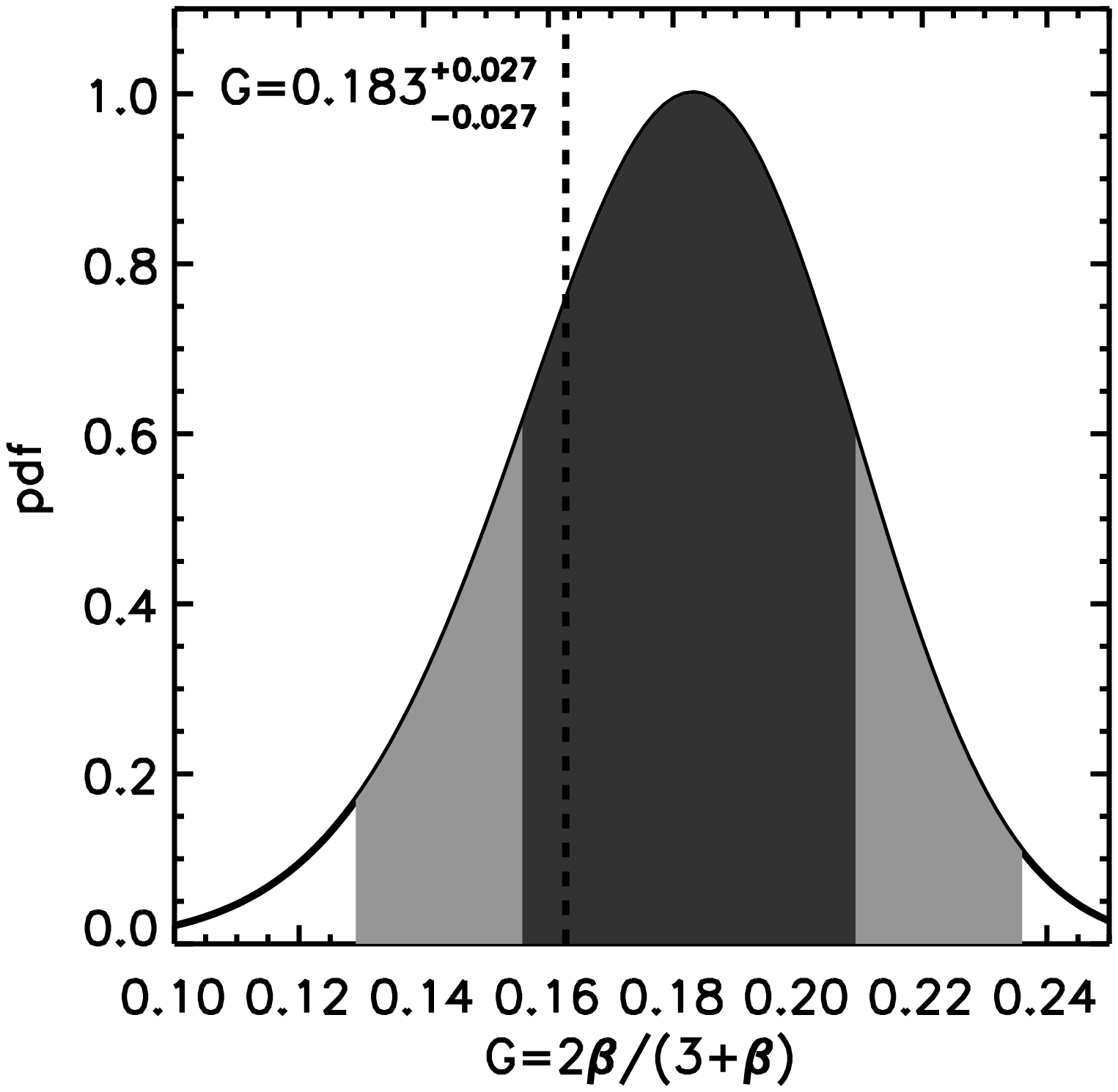}}
\caption{Likelihoods of the growth parameter $G$ for the measurements of the three sub-samples of voids: $r_v=25-30\Mpc$, $r_v=30-40\Mpc$ and $r_v=40-60\Mpc$ from the left to the right. Top panels are results using the linear model. Bottom panels are results from using the quasi-linear model with two free parameters $[\sigma_0, G]$, where $\sigma_0$ is the amplitudes of the velocity dispersion profile taken to follow an error function form, i.e. $\sigma_{\rm v}(r)=\sigma_0 {\rm erf}(r/r_{\rm v})$. The $\sigma_0$ parameters are marginalised.}
\label{subsample}
\end{center}
\end{figure*}
 
Since large voids have larger volume and contain more haloes than
small ones, they are expected to have higher signal-to-noise per
void. We try weighting voids according to their volume in our
analysis, finding that the error budget is reduced by $\approx 10\%$
and the peak of the likelihood is shifted to the left by approximately
8\%, which brings in better agreement (within 1$\sigma$) with
the fiducial value.  The precision of the measurement of the growth
factor $G$ is about 8\% for the volume of 3(Gpc/$h$)$^3$, as shown by
the likelihood function computed from $\exp({-\chi^2/2})$ on the
bottom-left panel of Fig.~\ref{Fig:measurement}. This also translates
into a 9\% constraint for $\beta$ as the central value is slightly
lower than the case without weighting.
 
We find that including voids with radii smaller than 25$\Mpc$ does not
help to reduce the error significantly, given that the number of voids
is significantly increased, but inclusion of these voids introduces a bias in the best-fit
value of $\beta$. This is possibly because small voids are more likely
to be sub-voids: these essentially sample part of
the volume that has already been used, thus adding no more information
other than introducing covariance. In particular, we find that these small
voids are more likely to be void-in-cloud objects, i.e. voids that
live in overdense environments. Such systems tend to have
over-compensated profiles, which are more complicated to model, with
non-linear effects associated with the
over-compensated wall. Indeed, we have found that there is a scale-dependent
ratio between the real-space void-halo cross-correlation $\xi_{\rm vh}$
and the void-mass cross-correlation $\xi_{\rm vm}$ when smaller voids are included. 
It may also be that, for such small voids, neglect of their correlated
velocity field is not justified in the way that applies for the
larger voids (see the discussion in Section~\ref{Sec:estimator}).
 
For $r>0.5r_{\rm v}$, we fit our measurement with the quasi-linear model
shown in Eq.~(\ref{FisherModel}). Void profiles are taken from the measured monopoles and the radial velocity profiles are inferred from linear theory using the monopoles. Motivated
by the simulation results shown in Figs 1-3, we adopt an error function form for
the velocity dispersion profile and allow its amplitude to vary,
i.e. $\sigma_{\rm v}(r)=\sigma_0\, {\rm erf}(r/r_{v})$, where $\sigma_0$ is a
free parameter. We compute $\chi^2$s in the [G, $\sigma_0$] 2D
parameter space, and the resulting joint constraints on
$\sigma_0$ and $G$ are shown on the top-left
panel of Fig.~\ref{Fig:measurement_Dispersion}. The fiducial values
are found to be consistent with the measurement within
$1\sigma$. There is, as expected, a degree of correlation between
$G$ and $\sigma_0$, so that the conditional error on $G$ at fixed $\sigma_0$ is
smaller than the true error on $G$, marginalized over 
$\sigma_0$; this yields the constraint on $G$ shown in the top-right panel of
Fig.~\ref{Fig:measurement_Dispersion}. However, because we are now able to
include data at smaller $r$, the error budget is a factor of
two smaller than when using the linear model with $r > r_{\rm v}$. Therefore, our
precision on $\beta$ in this case is $\approx 5\%$. The green
curve at the bottom panel of the figure shows our best-fit model, which
matches the data very
well throughout the range of scales shown. The impact of velocity
dispersion is indeed only important at $r<r_{\rm v}$. 
So $\sigma_0$ is constrained mainly by small scales, 
while $\beta$ is sensitive to the clustering throughout the whole range of scales; 
this explains why $\sigma_0$ and $\beta$ are not completely degenerate. 
With this model, we have again found that including smaller voids does not help to
tighten the constraint for the growth.
However, including these voids can have the effect of biasing
the recovered growth rate above the fiducial value, by as much
as 20\%. We therefore recommend that practical analyses should be
restricted to voids with $r_{\rm v}>25\Mpc$.

To check that the success of our measurements for the growth parameter
is robust, we split the void population into three sub-samples
covering different ranges of void radius, i.e. $r_v=25-30\Mpc$,
$r_v=30-40\Mpc$ and $r_v=40-60\Mpc$. We then repeat the analysis for
each with both the linear model and the quasi-linear model. The
likelihoods for the growth parameter $G$ (after marginalising over the
velocity dispersion parameter for the quasi-linear case) are shown in
Fig.~\ref{subsample}. All results are consistent with the fiducial
values within $2\sigma$ and the agreement for the results using the
quasi-linear model are slightly better. Note that the best-fit values
of $\sigma_0$ are also different for each of the three sub-samples.
This explains why the overall agreement here is better than when
fitting the whole sample assuming a universal value of $\sigma_0$
(Fig.~\ref{Fig:measurement_Dispersion}). Thus, this analysis of
sub-samples not only verifies the success of our method for measuring
the growth parameter, but also shows that it is better to split the
void sample into sub-samples of different radius when including a
velocity dispersion in the fitting.

Note that our measurement requires no prior knowledge of the void profile,
and works with quantities
that we can directly measure from the data. The
success of the measurement is guaranteed by the good agreement between
the recovered monopole from the redshift space correlation function
with the true answer, as shown by the overlap between the red dots and
the red dashed curves in the bottom-left panels of
Figs~\ref{Fig:LinearModel1}, \ref{Fig:LinearModel2} and
\ref{Fig:LinearModel3}. The measurement of these moments for individual
voids is very noisy as each void has its intrinsic
configuration, but by averaging measurements over all voids
in our sample, we are able to beat down the noise and recover
the monopole and quadrupoles without any obvious bias
(e.g. bottom-left panel of
Fig.~\ref{Fig:measurement}). We have also tested that the best-fit
results remains unchanged when we bin our voids in groups of 10 or 20,
conduct the measurement for each void composite and take the average.

The simplicity of the linear modelling of void-galaxy cross-correlations
contrasts with the
modelling of the redshift-space galaxy autocorrelation function, which
becomes difficult even at relatively large scales.
For example, recent measurements of the growth rate
using the galaxy autocorrelation function \citep{Samushia2014,Howlett2015} restricted
their analyses to scales greater than 24$\Mpc$ and 25$\Mpc$ respectively.
This is the same as the minimal void radius in our analysis, but 
the autocorrelation analysis requires careful modelling of 
velocity dispersion even to reach these scales, whereas we
have seen that it is unimportant for voids.
Adding the dispersion
term in the quasi-linear model allows us to use information
down to the much smaller scales of
$r=0.5r_{\rm v}$, or $12\Mpc$ for the void sample where the results are robust.

\subsection{Comparison to Hamaus et al.}
During the preparation of this manuscript, two papers on the same
topic were released \citep{Hamaus2015, Hamaus2016} [see also
    \cite{Mao2016} and \cite{Shoji2012} where the impact of
    redshift-space distortion on void ellipticity is studied in the
    latter].  Our work is distinct in terms of void definition,
methodology of measuring the growth as well as the prior assumptions
in the model fitting. \citet{Hamaus2015} and \citet{Hamaus2016} use
{\sc vide} \citep{Sutter2015}, which is based on the watershed
algorithm of {\sc zobov} \citep{Neyrinck2008} for HOD galaxies. We
employ the spherical underdensity void-finding algorithm for haloes
\citep{Padilla2005, Cai2015}. \citet{Hamaus2015} and
\citet{Hamaus2016} assumed a 4-parameter model for the void profile
(which largely depends on the void finding algorithm) and marginalised
over this model. In our methodology, no prior assumption about the
void profiles is needed, as we have shown that we can measure the void
profile from the monopole of the data. Our analysis pipeline is in
principle applicable to voids defined from any void-finding
algorithm. \citet{Hamaus2015} assumed a Lorentzian distribution for
the velocity dispersion, and fixed the width of the Lorentzian
function (using the best fit value from their simulations) prior to
their MCMC fitting. In our linear analysis, no model assumption for
the velocity is needed; and in our quasi-linear analysis, we only
assume a functional form for the dispersion profile without fixing
anything further. \citet{Hamaus2015} and \citet{Hamaus2016} have
  to bin their voids before fitting, which may come from the
  restriction that they have to assume a model for the void profile,
  and the model may not be able to describe individual profile. Our
  analysis is flexible as we have shown that binning our voids in sets
  of 10, 20 or not binning at all yields the same results.
\citet{Hamaus2015} included the distance distortion parameter related
to the AP test for their fitting.  We restrict our analysis to a
single free parameter of the growth for the purpose of understanding
how the configuration of voids are affected by peculiar velocity.

Perhaps more importantly, the results from
\citet{Hamaus2015} for the growth appear to be dominated by
systematics and are strongly biased from their fiducial values \footnote{The fiducial value of $\beta$ in \citet{Hamaus2015} at $z=0.5$ was incorrectly mentioned in their text. 
If the correct one was used, it is within $2\sigma$ of the best-fit value of $\beta$, but their best-fit value of the AP parameter lies outside the $2 \sigma$ contour.},
whereas our measurement for the growth displays no significant bias
within our chosen range of void sizes. We suspect that a possible
reason for the bias in the results of \citet{Hamaus2015}
is that they have used all data including the very
interior of voids and in some cases they have used voids with radii smaller than 10$\Mpc$. 
We have excluded this region ($r<0.5 r_{\rm v}$) from our fitting and restrict our analysis to using voids with $r_{\rm v}>25 \Mpc$. 
It is likely that for very small voids, or at void centres, structure formation is highly non-linear so that the quasi-linear model does not work there.

Finally, our results are also noticeably different from those of
\citet{Hamaus2016} for the best constraint on the growth parameter.
Using voids from a volume of $3.5(h^{-1}{\rm Gpc})^3$ from the SDSS DR11 CMASS
galaxy sample, they report constraints on $\beta$ at the 20\%
level \footnote{They have an additional parameter of $\Omega_m$, but
  as shown by their Fig. 3, even with $\Omega_m$ fixed at the fiducial
  value, the constraint on $\beta$ would remain similar. }. With a
similar volume of $3(h^{-1}{\rm Gpc})^3$ from simulation, our best constraint
on $\beta$ is a few times tighter, i.e. at the $5\%$ level.

\section{conclusions and discussion}
We have derived models for the void-halo or
void-galaxy correlation function in redshift space as well as their
Fourier space versions. 
We show that the linear-theory void-halo correlation function contains only monopole
and quadrupole terms, which allows us to write down an estimator for
the fluctuation growth rate based on the ratio of quadrupole and
monopole. We then test the estimator for voids found using halo density
fields from numerical simulation. We are able to extract the monopole from
the correlation function in an unbiased fashion at all scales of
interest, including the interior of voids. 
No prior knowledge of the void profile is needed, as it can be measured
from the data. 
This approach is to be contrasted with the work of
\cite{Ceccarelli2013} and \cite{Paz2013}, who used SDSS
voids to estimate the mean radial
velocity profile for voids, whereas we are able to obtain
this from linear theory.

The ratio between monopole and quadrupole provides an
unbiased estimates for the growth parameter $\beta=f/b$ at $r>r_{\rm v}$.
For an effective volume of
$3(h^{-1}{\rm Gpc})^3$, $\beta$ is constrained in this way with 9\% precision. 
Extending the model to allow for a dispersion of random velocities
allows us to apply the quasi-linear model at
smaller scales $r>0.5 r_{\rm v}$ and tighten the constraint for $\beta$
by a factor of two.  In principle, our estimator is applicable to any
void finding algorithm, since we have made no assumption beyond
linear theory and have no specific requirement for the void
population. We find that including scales below $r<0.5 r_{\rm v}$ can bias the results. This is probably because the centres of voids 
are evolving in a highly non-linear fashion and so it is beyond the description of the quasi-linear model.
We have noted that this lack of bias contrasts
with the results of the recent paper by \cite{Hamaus2015}, and
we have highlighted some methodological differences between our studies
that may account for this rather different outcome.

Our models allow us to understand
the complex distortion pattern of voids generated by peculiar velocities. We
find that voids may appear elongated or flattened along the line of
sight at different radii: this can be explained by the
competing amplitudes of the local density, radial velocity profiles
and their gradients, with the latter two being determined by the
cumulative density profile of voids in linear theory. Velocity
dispersion causes a slight elongation along the LOS for the correlation
function, which counters the flattening effect caused by the
streaming motion in the interior of the void and sometimes reverses it,
causing the sign of the quadrupole to flip. However, 
the effect of a random velocity dispersion is
usually negligible outside the void radius. Thus the dispersion
patten for voids is complex, and the key element that governs the void
distortion pattern at $r>r_{\rm v}$ is mostly the void profile. In light of
this, the distortion pattern for voids-in-clouds and voids-in-voids
are expected to be different. The picture will change when
there is strong non-linearity, but we have demonstrated that for large
voids with radii greater than 25$\Mpc$, the quasi-linear model works
very well.

The distortion patterns of voids in redshift space is more
    complex and is distinct from redshift-space distortions in
    overdensities.  For overdensities, infall motions cause flattening
    on large scales and random velocity dispersions induce elongation
    along the line of sight on small scales.

Our study implies that Alcock-Paczynski (AP) measurements using voids
will be affected by peculiar velocity distortions in a complex
manner. Assuming the correct cosmology, the dimension of voids along
the LOS may appear greater or smaller than that in the transverse
direction. It depends on the void population, and for the same void
population there is also a radial dependence. Ultimately, knowing the
void profile is the key to understanding the impact of peculiar
velocities on the apparent axial ratios of voids and on the resulting AP
measurement. Fortunately, we have shown that the redshift-space void
profile can be successfully extracted from the void-halo correlation
function. This may provide the key information to correct for the AP
measurement using voids.

A striking conclusion of this study is that 
the void-galaxy correlation function may be able to
provide a high precision on the growth rate (5\% for $\beta$ for a
volume of $3(h^{-1}{\rm Gpc})^3$, comparable to that obtained from using the LRG
autocorrelation function -- e.g. \cite{Samushia2014} found a precision of 10\% from a volume
of $2(h^{-1}{\rm Gpc})^3$. 
This success depends on
the applicability of the quasi-linear model at small scales $r \approx 12\Mpc$;
this is easier to achieve using voids, as 
the signal from extreme overdensities is then removed, and these are hardest to model. 
Admittedly, adding the AP test parameter may degrade the constraining power for the growth, but perhaps only very mildly 
[see for example, \citep{Hamaus2016}]. We leave this investigation for a future work. 
The next step will be to see if these virtues are maintained when using
mock data that accurately match real galaxy catalogues. If this is
indeed the case, then we will have not only a tool for assessing the
robustness of growth-rate measurements, but also a unique probe of deviations
from standard Einstein gravity, which are expected to reveal themselves
most strongly in low-density environments.

\section*{Acknowledgements}
We thank Baojiu Li for providing the N-body simulation used for this
study and for his useful comments. We thank Nico Hamaus, Mark Neyrinck and Istvan Szapudi for useful comments on
an earlier version of this paper. YC was supported during this work
by funding from an STFC Consolidated Grant. YC and JAP were supported by the European Research Council under grant number 670193.
ANT thanks the Royal Society for support from a
Wolfson Research Merit Award. NP was supported by BASAL PFB-06 CATA and Fondecyt 1150300.  NP used the Geryon cluster at the AIUC for the void finding algorithm.

\end{document}